\documentclass[11pt, reqno]{amsart}
\usepackage{amsfonts,latexsym}
\usepackage{amsmath}
\usepackage{amscd}
\usepackage{float,amsmath,amssymb,mathrsfs,bm,multirow,graphics}
\usepackage[dvips]{graphicx}
\usepackage[percent]{overpic}

\newcommand{\nequation}{\setcounter{equation}{0}}
\renewcommand{\theequation}{\mbox{\arabic{section}.\arabic{equation}}}

\newcommand{\C}{{\Bbb C}}
\newcommand{\Z}{{\Bbb Z}}

\newcommand{\proofbegin}{\noindent{\it Proof.\quad}}
\newcommand{\proofend}{\hfill$\Box$\bigskip}

\newcommand{\lot}{\text{\upshape lower order terms}}
\newcommand{\tr}{\text{\upshape tr\,}}
\newcommand{\res}{\text{\upshape Res\,}}
\newcommand{\diag}{\text{\upshape diag\,}}
\newcommand{\re}{\text{\upshape Re\,}}
\newcommand{\im}{\text{\upshape Im\,}}

\newtheorem{theorem}{Theorem}[section]
\newtheorem{proposition}[theorem]{Proposition}
\newtheorem{lemma}[theorem]{Lemma}

\newtheorem{assumption}[theorem]{Assumption}
\newtheorem{remark}[theorem]{Remark}

\newtheorem{figuretext}{Figure}

\input epsf
\date{\today}
\title[Initial-boundary value problems for integrable]
{Initial-boundary value problems for integrable evolution equations with $3 \times 3$ Lax pairs}

\author{Jonatan Lenells}
\address{Department of Mathematics, Baylor University, One Bear Place \#97328, Waco, TX 76798, USA.}
\email{Jonatan\_Lenells@baylor.edu}

\begin{document}

\begin{abstract} 
\noindent
We present an approach for analyzing initial-boundary value problems for integrable equations whose Lax pairs involve $3 \times 3$ matrices. Whereas initial value problems for integrable equations can be analyzed by means of the classical Inverse Scattering Transform (IST), the presence of a boundary presents new challenges. Over the last fifteen years, an extension of the IST formalism developed by Fokas and his collaborators has been successful in analyzing boundary value problems for several of the most important integrable equations with $2 \times 2$ Lax pairs, such as the Korteweg-de Vries, the nonlinear Schr\"odinger, and the sine-Gordon equations. In this paper, we extend these ideas to the case of equations with Lax pairs involving $3 \times 3$ matrices. 
\end{abstract}

\maketitle




\section{Introduction}\nequation
Several of the most important PDEs in mathematics and physics are integrable. Integrable PDEs can be analyzed by means of the Inverse Scattering Transform (IST) formalism, whose introduction was one of the most important developments in the theory of nonlinear PDEs in the 20th century. Until the 1990's the IST methodology was pursued almost entirely for pure initial value problems. However, in many laboratory and field situations, the wave motion is initiated by what corresponds to the imposition of boundary conditions rather than initial conditions. This naturally leads to the formulation of an initial-boundary value (IBV) problem instead of a pure initial value problem. 

In 1997, Fokas announced a new unified approach for the analysis of IBV problems for linear and nonlinear integrable PDEs \cite{F1997, F2002} (see also \cite{Fbook}). The Fokas method provides a generalization of the IST formalism from initial value to IBV problems, and over the last fifteen years, this method has been used to analyze boundary value problems for several of the most important integrable equations with $2 \times 2$ Lax pairs, such as the Korteweg-de Vries, the nonlinear Schr\"odinger, the sine-Gordon, and the stationary axisymmetric Einstein equations, see e.g. \cite{BFS2006, BS2003, BS2004, F2002, Fbook, FIS2005, Kamvissis, Lholedisk, LFernst, MK2006, P2005}. Just like the IST on the line, the unified method yields an expression for the solution $u(x,t)$ of an IBV problem in terms of the solution of a Riemann-Hilbert (RH) problem. In particular, the asymptotic behavior of $u(x,t)$ can be analyzed in an effective way by using this RH problem and by employing the nonlinear version of the steepest descent method introduced by Deift and Zhou \cite{DZ1993}. 

The purpose of this paper is to develop a methodology for analyzing IBV problems for integrable evolution equations with Lax pairs involving $3 \times 3$ matrices. 
Although the transition from $2\times 2$ to $3 \times 3$ matrix Lax pairs involves a number of novelties, the two main steps of the method of \cite{F1997, F2002} remain the same: (a) Construct an integral representation of the solution characterized via a matrix RH problem formulated in the complex $k$-plane, where $k$ denotes the spectral parameter of the Lax pair. This representation involves, in general, some unknown boundary values, thus the solution formula is not yet effective. (b) Characterize the unknown boundary values by analyzing the so-called global relation. In general, the characterization of the unknown boundary values involves the solution of a nonlinear problem. However, for so-called {\it linearizable} boundary conditions, this problem can be by-passed since the unknown boundary values can be eliminated using only algebraic manipulations.

In this paper, we will show how steps (a) and (b) can be implemented for a prototypical example of an equation with a $3 \times 3$ Lax pair. We expect that a similar analysis will apply also to other integrable equations with $3 \times 3$ Lax pairs, both on the half-line and on the interval, although in certain cases additional technical difficulties will arise. For example, for the Boussinesq equation the behavior of the eigenfunctions near the point $k=0$ requires special attention.\footnote{A similar situation arises for KdV for which the eigenfunctions have poles at $k=0$, see \cite{FI1994}.} Physically relevant equations with $3 \times 3$ Lax pairs include the Boussinesq, Degasperis-Procesi \cite{DP1999}, Kaup-Kupershmidt \cite{K1980}, Sasa-Satsuma \cite{SS1991}, Sawada-Kotera \cite{SK1974}, two-component vector nonlinear Schr\"odinger \cite{M1974}, and $3$-wave resonant interaction \cite{ZM1973} equations.

\subsection{The transition from $2\times 2$ to $3 \times 3$ Lax pairs}
Let us comment on some of the implications of the transition from $2\times 2$ to $3 \times 3$ Lax pairs. The implementation of the step (a) mentioned earlier in the case of $2 \times 2$ matrix Lax pairs is achieved by introducing eigenfunctions $\{\mu_j(x,t,k)\}$ which are defined by integration from the `corners' of the physical domain, see \cite{Fbook}. The column vectors of the $\mu_j$'s are bounded and analytic in different sectors of the complex $k$-plane and these column vectors are easily combined into a sectionally analytic function suitable for the formulation of a RH problem. However, for a Lax pair involving $3\times 3$-matrices, the bounded and analytic eigenfunctions suitable for the formulation of a RH problem involve rather complicated combinations of the entries of the $\mu_j$'s. Moreover, because of the limited domains of boundedness of the $\mu_j$'s, the boundedness properties of these combinations are not evident. We will therefore use a different approach for finding these combinations and their boundedness domains. Instead of taking the $\mu_j$'s as our starting point, we will define analytic eigenfunctions, denoted by $\{M_n(x,t,k)\}$, via integral equations which involve integration from all three corners simultaneously. 
In the absence of bound states, this formulation is adequate. However, since the integral equations defining the eigenfunctions now are of Fredholm rather than Volterra type, there may exist points $\{k_j\}$, $k_j \in \C$, at which the $M_n$'s have singularities. In order to deal with these singularities (which are related to the existence of solitons), we will relate the $M_n$'s to the $\mu_j$'s by solving a matrix factorization problem.

Another difference in the implementation of step (a) is that the RH problem in the case of a $2Ê\times 2$ Lax pair splits the complex $k$-plane into $4$ sectors, whereas a larger number of sectors is in general required in the case of $3 \times 3$ Lax pairs, e.g. for our main example, the RH problem splits the complex $k$-plane into $12$ sectors.

The implementation of the step (b) mentioned earlier involves eliminating the unknown boundary values from the formulation of the RH problem. In the case of linearizable boundary conditions for equations with $2 \times 2$ matrices, this elimination is achieved by algebraic manipulation of the so-called global relation and the equations obtained from the global relation under certain transformations in the $k$-plane. As shown in section \ref{linearizablesec} below, similar ideas can be used to analyze linearizable boundary conditions in the case of $3 \times 3$ Lax pairs. However, in the $3 \times 3$ case the algebraic relations cannot always be directly used to eliminate the unknown boundary values from the definition of the jump matrices. Instead, we will first analytically extend the domain of definition of the jump matrix, before we utilize the algebraic relations. Finally, we perform another analytic continuation to find the expression for the jump matrix on the relevant contour. The analyticity domains of the involved matrices are just sufficient to allow for this approach.

Let us finally point out that some pioneering works on the theory of inverse scattering for initial-value problems for equations with $3 \times 3$ Lax pairs are \cite{BC1984, DTT1982, K1980}. Other examples of the use of piecewise analytic solutions in studying the inverse problems for $n \times n$ systems (particularly for $n = 4$) can be found in \cite{BS1998, BS2000, SS2000}.

\subsection{Organization of the paper}
Our main example is introduced in subsection \ref{mainexamplesubsec}. The spectral analysis of the associated Lax pair is performed in section \ref{specanalysissec}. In section \ref{RHsec}, we formulate the main RH problem, and this concludes the implementation of step (a) above. 

Step (b) is implemented in sections \ref{linearizablesec} and \ref{nonlinearizablesec}. 
In section \ref{linearizablesec}, we consider linearizable boundary conditions, whereas nonlinearizable boundary conditions are analyzed in section \ref{nonlinearizablesec}.
Section \ref{conclusionssec} contains some concluding remarks. In appendix \ref{Aapp}, we explain the relationship between the constructions of this paper and the formalism of \cite{F1997, F2002} for $2 \times 2$ Lax pairs. In appendix \ref{fredholmapp}, we use an extension of the standard Fredholm theory to study a set of integral equations.

\subsection{The main example}\label{mainexamplesubsec}
To be concrete, we will consider the system
\begin{align}\label{qrsystem}
\begin{cases}
iq_t + \frac{1}{\sqrt{3}}q_{xx} + 2irr_x = 0, \\
ir_t - \frac{1}{\sqrt{3}}r_{xx} + 2iqq_x = 0,
\end{cases}
\end{align}
where $q(x,t)$ and $r(x,t)$ are complex-valued functions of $(x,t) \in \Omega$, with $\Omega$ denoting the half-line domain
$$\Omega = \{0 < x < \infty, \; 0 < t < T\}$$ 
and $T>0$ being a fixed final time. This system is the compatibility condition of the $3 \times 3$ Lax pair
\begin{align}\label{psilaxV1V2}
\begin{cases}
\psi_x - kJ\psi = V_1\psi, \\
\psi_t + k^2J^2 \psi = V_2\psi,
\end{cases} \qquad J = \begin{pmatrix} 1 & 0 & 0 \\ 
0 & \omega^2 & 0 \\
0 & 0 & \omega
\end{pmatrix}, \qquad \omega = e^{\frac{2\pi i}{3}},
\end{align}
where $\psi(x,t,k)$ is a $3 \times 3$-matrix valued function, $k \in \C$ is the spectral parameter, and $\{V_1(x,t), V_2(x,t,k)\}$ are $3 \times 3$-matrix valued functions 
given by
\begin{align}
& V_1 =  \begin{pmatrix}
0 & q & r \\
 r & 0 & q \\
 q & r & 0
\end{pmatrix}, \qquad  V_2(x,t,k) = kV_2^{(1)}(x,t) + V_2^{(0)}(x,t), 
	\\ \nonumber
& V_2^{(1)} = \begin{pmatrix}
 0 & \omega  q & \omega ^2 r \\
 \omega  r & 0 & q \\
 \omega ^2 q & r & 0\end{pmatrix},
 \quad
 V_2^{(0)} = \frac{i}{\sqrt{3}}
\begin{pmatrix}
 0 & q_x & -r_x \\
 -r_x & 0 & q_x \\
 q_x & -r_x & 0\end{pmatrix}
 -
 \begin{pmatrix}
 0 & r^2 & q^2 \\
 q^2 & 0 & r^2 \\
 r^2 & q^2 & 0
\end{pmatrix}.
\end{align}
We will denote the initial data of (\ref{qrsystem}) by $\{q_0(x), r_0(x)\}$, while the Dirichlet and Neumann boundary values will be denoted by $\{g_0(t), h_0(t)\}$ and $\{g_1(t), h_1(t)\}$, respectively, i.e.
\begin{align}\nonumber
 & q(x,0) = q_0(x), \quad r(x,0) = r_0(x), \qquad 0 < x < \infty;
  	\\ \label{boundaryvalues}
 & q(0,t) = g_0(t), \quad r(0,t) = h_0(t), 
 	\\\nonumber
& q_x(0,t) = g_1(t), \quad r_x(0,t) = h_1(t), \qquad 0 < t < T.
\end{align}

\begin{remark}\upshape
1. We have chosen (\ref{qrsystem}) as our main example, because the Lax pair (\ref{psilaxV1V2}) is a natural $3\times 3$ generalization of the $2\times 2$ Lax pair for the nonlinear Schr\"odinger equation. Indeed, the $x$ and $t$ parts of the Lax pair for NLS involve the matrices $k\sigma_3$ and $k^2\sigma_3$, respectively, where $\sigma_3$ is the diagonal matrix whose entries are the second roots of unity, i.e. $\sigma_3 = \diag(1, -1)$. Analogously, the $x$ and $t$ parts of (\ref{psilaxV1V2}) involve the matrices $kJ$ and $k^2J^2$, where $J$ is the diagonal matrix whose entries are the third roots of unity.

2. When $r = \bar{q}$, the system (\ref{qrsystem}) reduces to the following integrable equation (cf. Eq. (1.5) in \cite{M1981}):
\begin{align}\label{qequation}
iq_t + \frac{1}{\sqrt{3}}q_{xx} + 2i\bar{q}\bar{q}_x = 0.
\end{align}
\end{remark}

\section{Spectral analysis}\nequation\label{specanalysissec}
Our goal in this section is to define sectionally analytic eigenfunctions of the Lax pair (\ref{psilaxV1V2}) which are suitable for the formulation of a Riemann-Hilbert problem. 

\subsection{The closed one-form}
Suppose that $q(x,t)$ and $r(x,t)$ are sufficiently smooth functions of $(x,t)$ in the half-line domain $\Omega$ which decay as $x \to \infty$.
Introducing $L(x,t,k)$ and $Z(x,t,k)$ by
\begin{align}\label{LZdef}
L = kJ + V_1, \qquad Z = -k^2J^2 + V_2,
\end{align}
we can write (\ref{psilaxV1V2}) as
\begin{align}\label{psilax}
\begin{cases}
\psi_x = L\psi, \\
\psi_t = Z \psi.
\end{cases}
\end{align}
The functions $L$ and $Z$ satisfy
$$\tr L = 0, \qquad \tr Z = 0,$$
as well as the $\Z_3$ symmetry
\begin{align}\label{Z3symm}
L(k) = \mathcal{A} L(\omega k)\mathcal{A}^{-1}, \quad
Z(k) = \mathcal{A} Z(\omega k)\mathcal{A}^{-1}, \quad
\mathcal{A} =  \begin{pmatrix}
0 & 1 & 0 \\
 0 & 0 & 1 \\
 1 & 0 & 0\end{pmatrix}.
\end{align}
In order to keep the notation concise, we define diagonal matrices $\mathcal{L}$ and $\mathcal{Z}$ by
$$\mathcal{L} = \lim_{x \to \infty} L = kJ, \qquad \mathcal{Z} =  \lim_{x \to \infty} Z = -k^2 J^2,$$
and denote the diagonal entries of these matrices by $\{l_j(k)\}_1^3$ andÊ$\{z_j(k)\}_1^3$:
$$\mathcal{L} = \diag(l_1, l_2, l_3), \qquad 
\mathcal{Z} = \diag(z_1, z_2, z_3).$$
Introducing a new eigenfunction $\mu(x,t,k)$ by
\begin{equation}\label{psimurelation}  
  \psi = \mu e^{\mathcal{L} x + \mathcal{Z} t},
\end{equation}
we find the Lax pair equations
\begin{equation}\label{mulax}
\begin{cases}
\mu_x - [\mathcal{L}, \mu] = V_1\mu, \\
\mu_t - [\mathcal{Z}, \mu] = V_2\mu.
\end{cases}
\end{equation}
Letting $\hat{\mathcal{L}}$ and $\hat{\mathcal{Z}}$ denote the operators which act on a $3 \times 3$ matrix $X$ by $\hat{\mathcal{L}}X = [\mathcal{L}, X]$ and $\hat{\mathcal{Z}}X = [\mathcal{Z}, X]$, the equations in (\ref{mulax}) can be written in differential form as
\begin{equation}\label{mulaxdiffform}
d\left(e^{-\hat{\mathcal{L}}x - \hat{\mathcal{Z}}t} \mu \right) = W,
\end{equation}
where $W(x,t,k)$ is the closed one-form defined by
\begin{equation}\label{Wdef}  
  W = e^{-\hat{\mathcal{L}}x - \hat{\mathcal{Z}}t}(V_1 dx + V_2 dt) \mu.
\end{equation}

\begin{remark}\upshape
In the case of equation (\ref{qequation}), the assumption $\bar{q} = r$ implies that the Lax pair possesses the following additional $\Z_2$-symmetry:
$$L(k) = \mathcal{B} \overline{L(\bar{k})} \mathcal{B}^{-1}, \qquad
Z(k) = \mathcal{B} \overline{Z(\bar{k})} \mathcal{B}^{-1}, \qquad
\mathcal{B} = \begin{pmatrix}
 1 & 0 & 0 \\
 0 & 0 & 1 \\
 0 & 1 & 0\end{pmatrix}.$$
\end{remark}

\subsection{The $M_n$'s}
\begin{figure}
\begin{center}
\medskip
 \begin{overpic}[width=.25\textwidth]{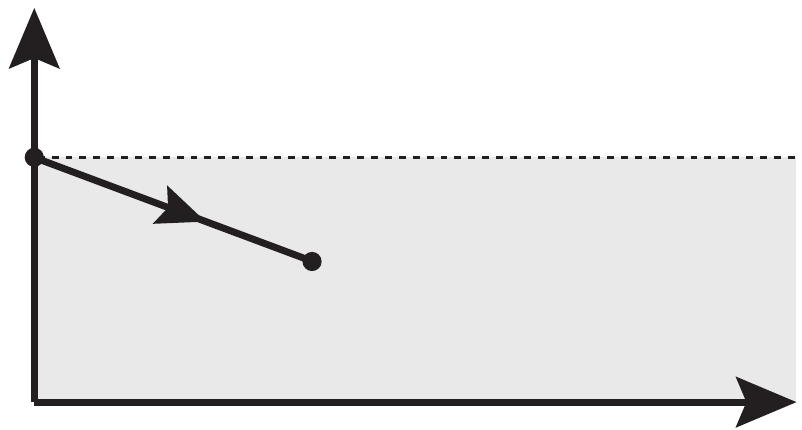}
 \put(-8,30){\small $T$}
 \put(43,17){\small $(x,t)$}
 \put(48,-10){\small $\gamma_1$}
   \end{overpic}
   \qquad\quad
  \begin{overpic}[width=.25\textwidth]{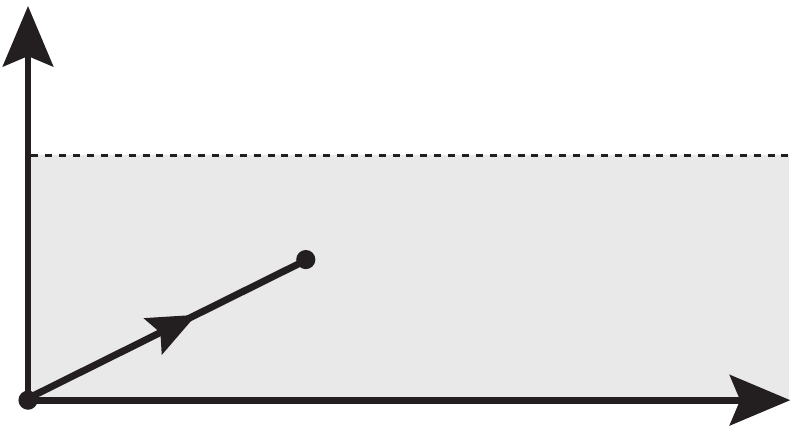}
    \put(-8,30){\small $T$}
    \put(43,18){\small $(x,t)$} 
 \put(48,-10){\small $\gamma_2$}
  \end{overpic} 
  \qquad \quad
  \begin{overpic}[width=.25\textwidth]{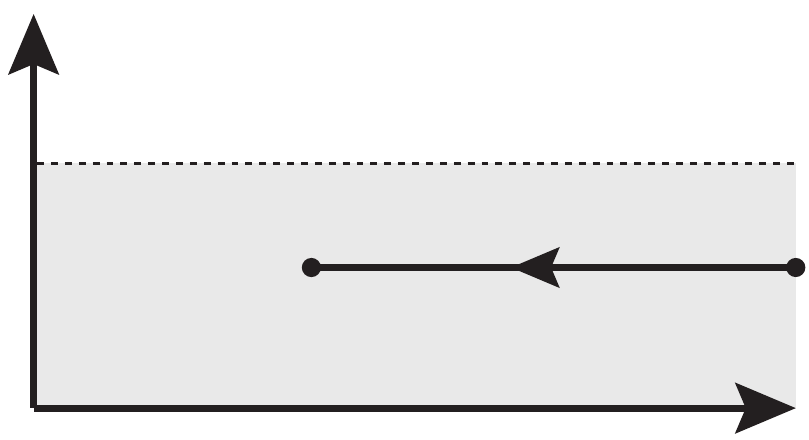}
     \put(-8,30){\small $T$}
    \put(13,18){\small $(x,t)$} 
 \put(48,-10){\small $\gamma_3$}
  \end{overpic}
  \bigskip
   \begin{figuretext}\label{mucontoursfig}
       The contours $\gamma_1$, $\gamma_2$, and $\gamma_3$ in the $(x, t)$-plane.
   \end{figuretext}
   \end{center}
\end{figure}

\begin{figure}
\begin{center}
\bigskip
 \begin{overpic}[width=.55\textwidth]{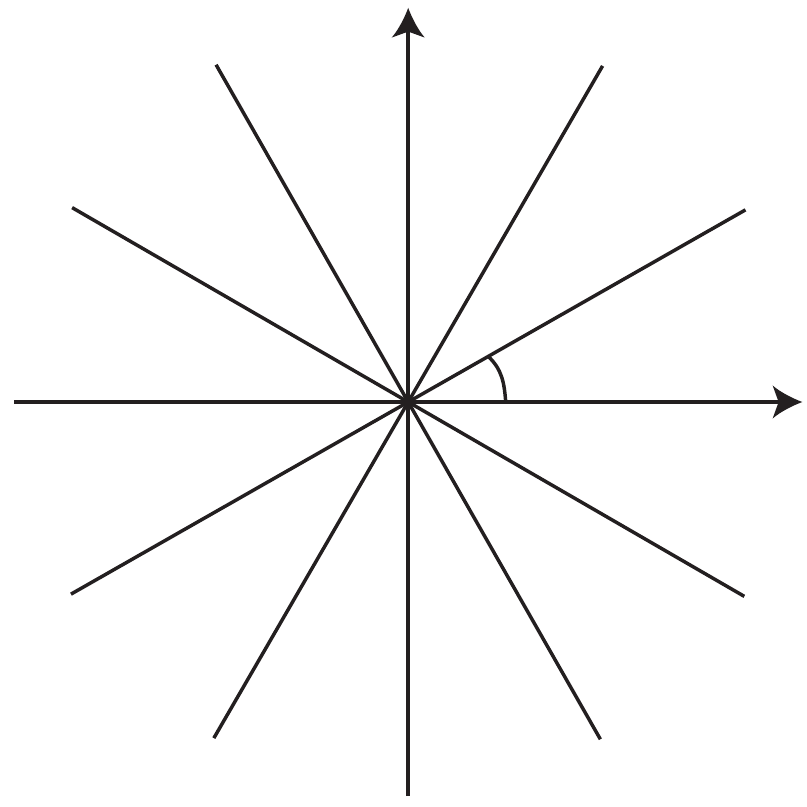}
 \put(65,53){\small $\pi/6$}
 \put(101,49){\small $\re k$}
 \put(46,101){\small $\im k$}
 \put(86,58){\small $D_1$}
 \put(75,75){\small $D_2$}
 \put(58,84){\small $D_3$}
 \put(38,84){\small $D_4$}
 \put(22,75){\small $D_5$}
 \put(12,58){\small $D_6$}
 \put(12,39){\small $D_7$}
 \put(20,20){\small $D_8$}
 \put(38,11){\small $D_9$}
 \put(58,11){\small $D_{10}$}
 \put(75,20){\small $D_{11}$}
 \put(86,39){\small $D_{12}$}
   \end{overpic}
   \begin{figuretext}\label{Dns.pdf}
      The sets $D_n$, $n = 1, \dots, 12$, which decompose the complex $k$-plane. 
      \end{figuretext}
   \end{center}
\end{figure}

Let $\{\gamma_j\}_1^3$ be contours in the $(x, t)$-plane which connect $(x_j, t_j)$ to $(x,t)$, where 
$$(x_1, t_1) = (0, T), \quad (x_2, t_2) = (0, 0), \quad (x_3, t_3) = (\infty, t).$$ 
We will assume that $T < \infty$, except for in section \ref{linearizablesec} where it will be assumed that $T = \infty$. 
Choosing the particular contours shown in Figure \ref{mucontoursfig}, we have the following inequalities on the contours:
\begin{align}\nonumber
\gamma_1: x - x' \geq 0,& \qquad t - t' \leq 0,
	\\ \label{contourinequalities}
\gamma_2: x - x' \geq 0,& \qquad t - t' \geq 0,
	\\ \nonumber
\gamma_3: x - x' \leq 0.&
\end{align}
For each $n = 1, \dots, 12$, define a solution $M_n(x,t,k)$ of (\ref{mulax}) by the following system of integral equations:
\begin{equation}\label{Mnintegraleq}
(M_n)_{ij}(x,t,k) = \delta_{ij} + \int_{\gamma_{ij}^n} \left(e^{\hat{\mathcal{L}}(k)x + \hat{\mathcal{Z}}(k)t} W_n(x',t',k)\right)_{ij}, \qquad k \in D_n, \quad i,j = 1, 2,3,
\end{equation}
where $W_n$ is given by (\ref{Wdef}) with $\mu$ replaced by $M_n$, the contours $\gamma^n_{ij}$, $n = 1, \dots, 12$, $i,j = 1, 2,3$, are defined by
 \begin{align} \label{gammaijnudef}
 \gamma_{ij}^n =  \begin{cases}
 \gamma_1  \quad \text{if} \quad \re  l_i(k) < \re  l_j(k) \;\; \text{and} \; \; \re  z_i(k) \geq \re  z_j(k)
	\\
\gamma_2  \quad \text{if} \quad \re  l_i(k) < \re  l_j(k)  \;\; \text{and} \; \;  \re  z_i(k) < \re  z_j(k)
	\\
\gamma_3  \quad \text{if} \quad \re  l_i(k) \geq \re  l_j(k)
	\\
\end{cases} \quad \text{for} \quad k \in D_n,
\end{align}
and $\{D_n\}_1^{12}$ denote twelve open, pairwisely disjoint subsets of the complex $k$-plane defined by (see Figure \ref{Dns.pdf})
\begin{align*}
&D_1 = \{k \in \C\,|\, \re l_3 < \re l_2 < \re l_1 \text{  and  } 
\re z_1 < \re z_3 < \re z_2 \},
	\\
&D_2 = \{k \in \C\,|\, \re l_3 < \re l_2 < \re l_1 \text{  and  } 
\re z_3 < \re z_1 < \re z_2 \},
	\\
&D_3 = \{k \in \C\,|\, \re l_3 < \re l_1 < \re l_2 \text{  and  } 
\re z_3 < \re z_2 < \re z_1 \},
	\\
&D_4 = \{k \in \C\,|\, \re l_3 < \re l_1 < \re l_2 \text{  and  } 
\re z_2 < \re z_3 < \re z_1 \},
	\\
&D_5 = \{k \in \C\,|\, \re l_1 < \re l_3 < \re l_2 \text{  and  } 
\re z_2 < \re z_1 < \re z_3 \},
	\\
&D_6 = \{k \in \C\,|\, \re l_1 < \re l_3 < \re l_2 \text{  and  } 
\re z_1 < \re z_2 < \re z_3 \},
	\\
&D_7 = \{k \in \C\,|\, \re l_1 < \re l_2 < \re l_3 \text{  and  } 
\re z_1 < \re z_3 < \re z_2 \},
	\\
&D_8 = \{k \in \C\,|\, \re l_1 < \re l_2 < \re l_3 \text{  and  } 
\re z_3 < \re z_1 < \re z_2 \},
	\\
&D_9 = \{k \in \C\,|\, \re l_2 < \re l_1 < \re l_3 \text{  and  } 
\re z_3 < \re z_2 < \re z_1 \},
	\\
&D_{10} = \{k \in \C\,|\, \re l_2 < \re l_1 < \re l_3 \text{  and  } 
\re z_2 < \re z_3 < \re z_1 \},
	\\
&D_{11} = \{k \in \C\,|\, \re l_2 < \re l_3 < \re l_1 \text{  and  } 
\re z_2 < \re z_1 < \re z_3 \},
	\\
&D_{12} = \{k \in \C\,|\, \re l_2 < \re l_3 < \re l_1 \text{  and  } 
\re z_1 < \re z_2 < \re z_3 \}.
\end{align*}
The following proposition ascertains that the $M_n$'s defined in this way have the properties required for the formulation of a RH problem.

\begin{proposition}\label{Mnprop}
For each $n = 1, \dots, 12$, the function $M_n(x,t,k)$ is well-defined by equation (\ref{Mnintegraleq}) for $k \in \bar{D}_n$ and $(x,t) \in \bar{\Omega}$. For any fixed point $(x,t)$, $M_n$ is bounded and analytic as a function of $k \in D_n$ away from a possible discrete set of singularities $\{k_j\}$ at which the Fredholm determinant vanishes. Moreover, $M_n$ admits a bounded and continuous extension to $\bar{D}_n$ and
\begin{align}\label{Mnatinfinity}
  M_n(x,t,k) = I + O(1/k), \qquad k \to \infty, \quad k \in D_n.
\end{align}  
\end{proposition}
\proofbegin
The boundedness and analyticity properties are established in appendix \ref{fredholmapp}; here we prove (\ref{Mnatinfinity}). Substituting the expansion
$$M = M^{(0)} + \frac{M^{(1)}}{k} + \frac{M^{(2)}}{k^2} + \cdots, \qquad k \to \infty, $$
into the $x$-part of the Lax pair (\ref{mulax}), the terms of $O(k)$ yield
$$[J, M^{(0)}] = 0 \quad \text{i.e.} \quad \text{$M^{(0)}$ is a diagonal matrix}.$$  
The diagonal terms of $O(1)$ imply that $M^{(0)}$ is a constant matrix. Evaluation of (\ref{Mnintegraleq})  at $(x_3, t_3)$ implies that $M^{(0)} = I$.
\proofend

\subsection{The jump matrices}
We define spectral functions $S_n(k)$, $n = 1, \dots, 12$, by
\begin{align}\label{Sndef}  
  & S_n(k) = M_n(0, 0, k), \qquad k \in D_n, \quad n = 1, \dots, 12.
\end{align}
Let $M$ denote the sectionally analytic function on the Riemann $k$-sphere which equals $M_n$ for $k \in D_n$. Then $M$ satisfies the jump conditions
\begin{equation}\label{MnMmrelation}  
  M_n = M_m J_{m,n}, \qquad k \in \bar{D}_n \cap \bar{D}_m, \qquad n, m = 1, \dots, 12, \quad n \neq m, 
\end{equation}
where the jump matrices $J_{m,n}(x, t, k)$ are defined by
\begin{equation}\label{Jmndef}
J_{m,n} = e^{\hat{\mathcal{L}}x + \hat{\mathcal{Z}}t}(S_m^{-1}S_n),\qquad n, m \in \{1, \dots, 12\}.
\end{equation}
Since the integral equations (\ref{Mnintegraleq}) that define $M_n(0, 0, k)$ only involve integration along the initial half-line $\{0 < x < \infty, \; t = 0\}$ and along the boundary $\{x = 0, \; 0 < t < T\}$, the $S_n$'s (and hence also the $J_{m,n}$'s) can be computed from the initial and boundary data alone. 
Thus, relation (\ref{MnMmrelation}) provides the jump condition for a RH problem, which, in the absence of singularities, can be used to reconstruct the solution $\{q(x,t), r(x,t)\}$ from the initial and boundary data. However, if the $M_n$'s have pole singularities at some points $\{k_j\}$, $k_j \in \C$, the RH problem needs to include the residue conditions at these points. 
For the purpose of determining the correct residue conditions (and also for the purposes of analyzing the linearizable and nonlinearizable boundary conditions in sections \ref{linearizablesec} and \ref{nonlinearizablesec}), it is convenient to introduce three eigenfunctions $\{\mu_j(x,t,k)\}_1^3$ in addition to the $M_n$'s.

\subsection{The $\mu_j$'s}\label{mujsubsec}
\begin{figure}
\begin{center}
\medskip
 \begin{overpic}[width=.44\textwidth]{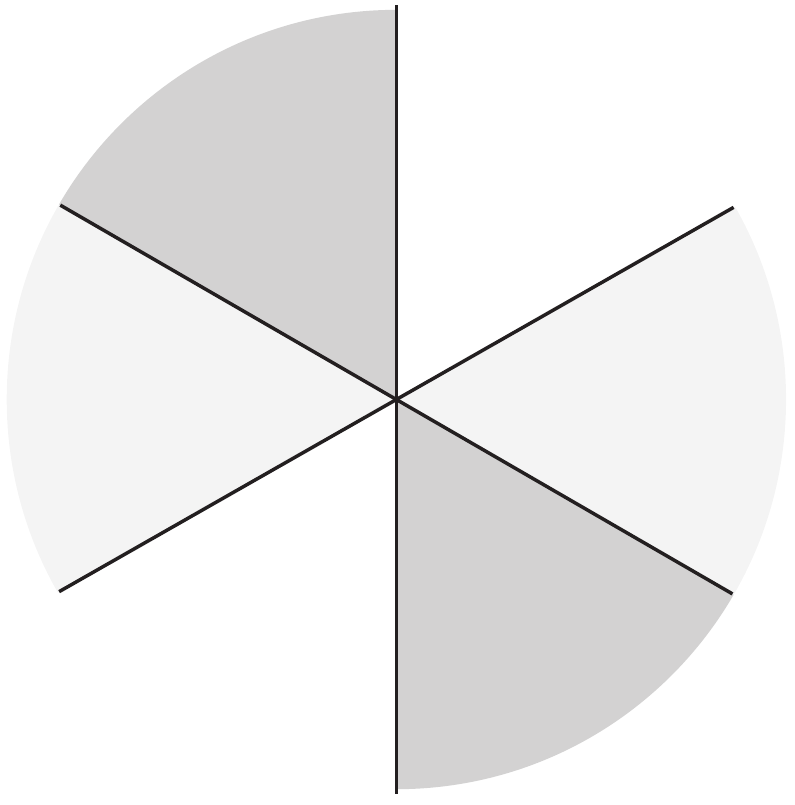}
 \put(32,22){\small $\mathcal{R}$}
 \put(75,49){\small $\omega \mathcal{R}$}
 \put(28,76){\small $\omega^2 \mathcal{R}$}
 \put(64,76){\small $\hat{\mathcal{R}}$}
 \put(15,49){\small $\omega \hat{\mathcal{R}}$}
 \put(61,22){\small $\omega^2 \hat{\mathcal{R}}$}
  \end{overpic}
  \qquad
   \begin{overpic}[width=.44\textwidth]{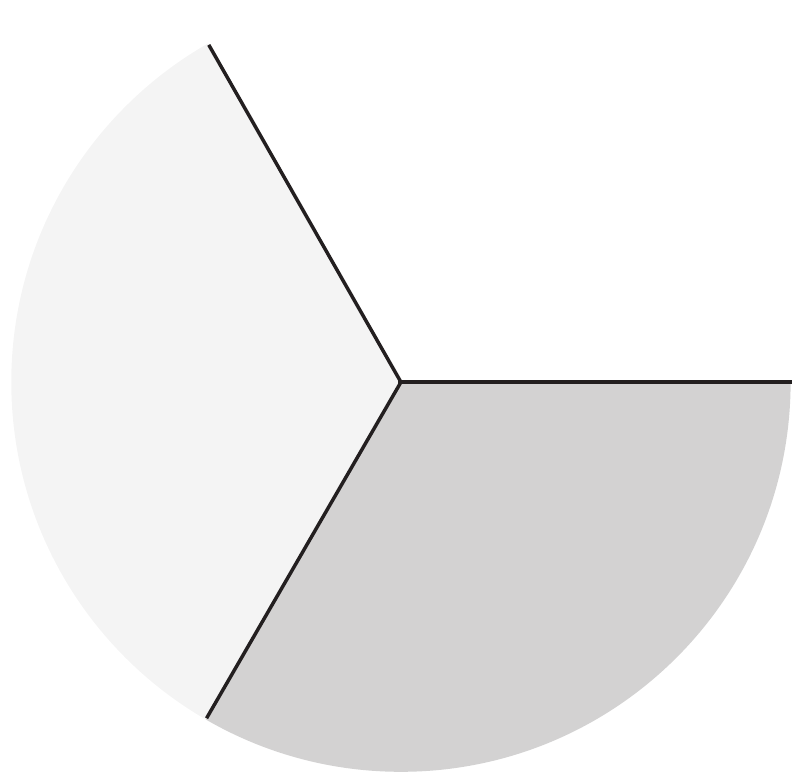}
 \put(65,72){\small $\mathcal{S}$}
 \put(20,49){\small $\omega \mathcal{S}$}
 \put(62,26){\small $\omega^2 \mathcal{S}$}
  \end{overpic}
  \\
  \vspace{.2cm}
   \begin{overpic}[width=.44\textwidth]{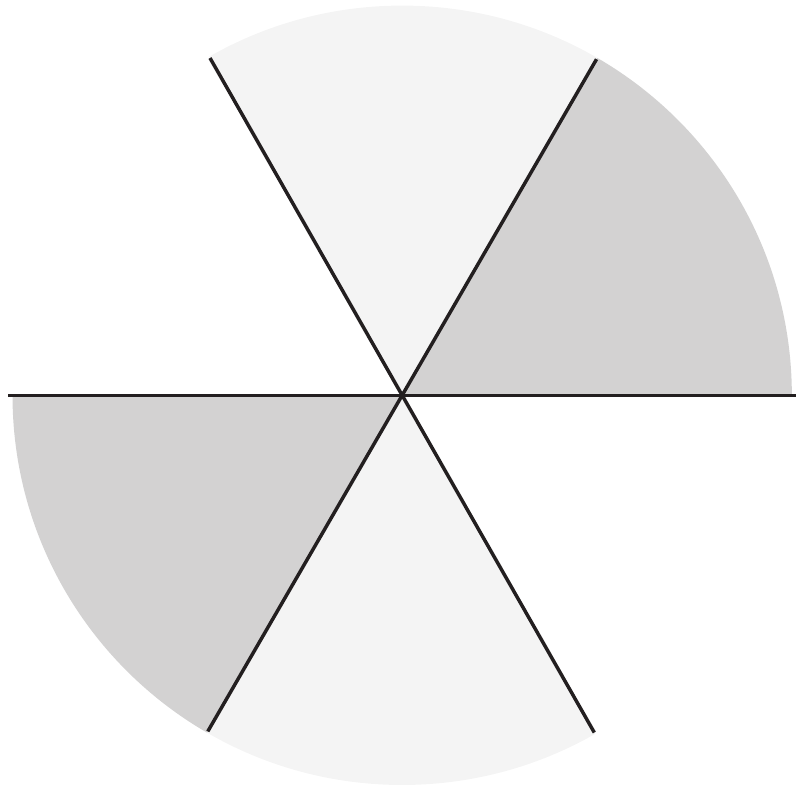}
 \put(20,64){\small $\mathcal{T}$}
 \put(47,18){\small $\omega \mathcal{T}$}
 \put(72,64){\small $\omega^2 \mathcal{T}$}
 \put(76,30){\small $\hat{\mathcal{T}}$}
 \put(46,78){\small $\omega \hat{\mathcal{T}}$}
  \put(18,31){\small $\omega^2 \hat{\mathcal{T}}$}
  \end{overpic}
 \qquad
  \begin{overpic}[width=.44\textwidth]{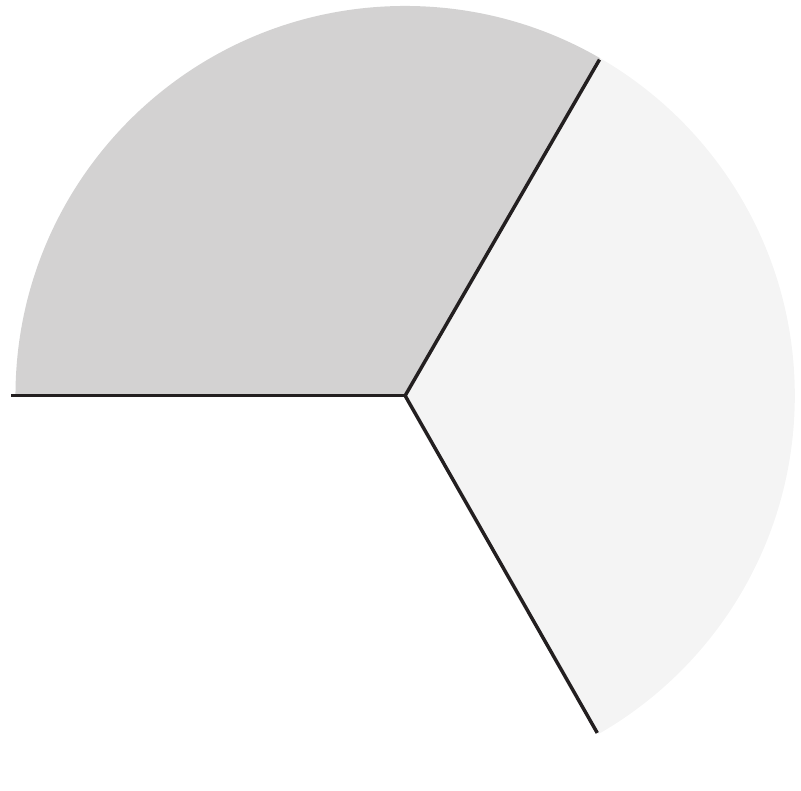}
 \put(32,22){\small $\hat{\mathcal{S}}$}
 \put(75,49){\small $\omega \hat{\mathcal{S}}$}
 \put(28,76){\small $\omega^2 \hat{\mathcal{S}}$}
   \end{overpic}
   \begin{figuretext}\label{RSTsetsfig}
      The decompositions of the complex $k$-plane defined by the sets $\mathcal{R}$, $\mathcal{S}$, $\mathcal{T}$ and their reflections $\hat{\mathcal{R}}$, $\hat{\mathcal{S}}$, $\hat{\mathcal{T}}$. 
      \end{figuretext}
   \end{center}
\end{figure}

We define three eigenfunctions $\{\mu_j\}_1^3$ of (\ref{mulax}) by the Volterra integral equations 
\begin{equation}\label{mujdef}
  \mu_j(x,t,k) = I +  \int_{\gamma_j} e^{\hat{\mathcal{L}}(k)x + \hat{\mathcal{Z}}(k)t} W_j(x',t',k), \qquad j = 1, 2,3,
\end{equation}
where $W_j$ is given by (\ref{Wdef}) with $\mu$ replaced with $\mu_j$. 
The third column of the matrix equation (\ref{mujdef}) involves the exponentials
$$e^{(l_1 - l_3)(x - x') + (z_1 - z_3)(t - t')}, \qquad e^{(l_2 - l_3)(x - x') + (z_2 - z_3)(t - t')}.$$ 
Using the inequalities in (\ref{contourinequalities}) it follows that these exponentials are bounded in the following regions of the complex $k$-plane:
\begin{align*}
\gamma_1: \{\re  l_1 < \re  l_3\} \cap \{\re  l_2 < \re  l_3\} &\cap \{\re  z_3 < \re  z_1\} \cap \{\re  z_3 < \re  z_2\},
	\\
\gamma_2: \{\re  l_1 < \re  l_3\} \cap \{\re  l_2 < \re  l_3\} &\cap \{\re  z_1 < \re  z_3\} \cap \{\re  z_2 < \re  z_3\},
	\\
\gamma_3: \{\re  l_3 < \re  l_1\} \cap \{\re  l_3 < \re  l_2\}&.
\end{align*}
Since the equations in (\ref{mujdef}) are Volterra integral equations, these boundedness properties imply that the third column vectors of $\mu_1$ and $\mu_3$ are bounded and analytic for $k \in \C$ such that $k$ belongs to $\mathcal{R}$ and $\mathcal{S}$, respectively,  where the sets $\mathcal{R}$ and $\mathcal{S}$ are defined by (see Figure \ref{RSTsetsfig})
$$\mathcal{R} = D_8 \cup D_9, \qquad \mathcal{S} = D_1 \cup D_2 \cup D_3 \cup D_{4}.$$ 
Similar conditions are valid for the other column vectors. Thus,
\begin{align*}
& \mu_1(x,t,k) \; \text{is bounded for} \; k \in (\omega \mathcal{R}, \omega^2 \mathcal{R}, \mathcal{R}),
	\\
& \mu_3(x,t,k)\; \text{is bounded for} \; k \in (\omega \mathcal{S}, \omega^2 \mathcal{S}, \mathcal{S}),
\end{align*}
where $k \in (\omega \mathcal{R}, \omega^2 \mathcal{R}, \mathcal{R})$ indicates that the first, second, and third columns of the equation are valid for $k$ in the sets $\omega \mathcal{R}$, $\omega^2\mathcal{R}$, and $\mathcal{R}$, respectively. On the other hand, the corresponding sets where the columns of $\mu_2$ are bounded are all empty.

Note that $\mu_1$ and $\mu_2$ are entire functions of $k$, whereas the first, second, and third columns of $\mu_3$, whose definitions involve integration from Ê$x = \infty$, are defined only for $k \in (\omega \mathcal{S}, \omega^2 \mathcal{S}, \mathcal{S})$.
Moreover, as $k \to \infty$,
\begin{align*}
& \mu_1(x, t, k) = I + O(1/k), \qquad k \to \infty, \quad k \in (\omega \mathcal{R}, \omega^2 \mathcal{R}, \mathcal{R}),
	\\
& \mu_3(x,t,k) = I + O(1/k), \qquad  k \to \infty, \quad k \in (\omega \mathcal{S}, \omega^2 \mathcal{S}, \mathcal{S}).
\end{align*}
For $x = 0$, $\mu_1$ and $\mu_2$ have the following enlarged domains of boundedness:
\begin{align*}
& \mu_1(0,t,k) \; \text{is bounded for} \; k \in (\omega\mathcal{R} \cup \omega \hat{\mathcal{R}}, \omega^2\mathcal{R} \cup \omega^2 \hat{\mathcal{R}}, 
\mathcal{R} \cup \hat{\mathcal{R}}),
	\\
& \mu_2(0,t,k) \; \text{is bounded for} \; k \in (\omega\mathcal{T} \cup \omega \hat{\mathcal{T}}, \omega^2\mathcal{T} \cup \omega^2 \hat{\mathcal{T}}, 
\mathcal{T} \cup \hat{\mathcal{T}}),
\end{align*}
where $\mathcal{T} = D_5 \cup D_6$ and, for a set $U \subset \C$, $\hat{U}$ denotes the reflection of $U$ with respect to the origin, i.e. $\hat{U} = \{k \in \C \, | \, -k \in U\}$; in particular, $\hat{\mathcal{R}} = D_2 \cup D_3$ andÊ $\hat{\mathcal{T}} = D_{11}\cup D_{12}$.

\begin{remark}\upshape
We have defined two sets of eigenfunctions: $\{M_n\}_1^{12}$ and $\{\mu_j\}_1^3$. The unified approach of \cite{F1997} for Lax pairs involving $2 \times 2$-matrices also implicitly utilizes two types of eigenfunctions: the $\mu_j$'s are used for the spectral analysis, whereas the RH problem is formulated in terms of another set of eigenfunctions; our $M_n$'s are the analogs of this latter set of eigenfunctions, see appendix \ref{Aapp}.
\end{remark}

\subsection{The minors of the eigenfunctions}
We will also need the analyticity and boundedness properties of the minors of the matrices $\mu_j(x,t,k)$, $j = 1,2,3$. To this end, we recall that the cofactor matrix $X^A$ of a $3 \times 3$ matrix $X$ is defined by
$$X^{A} = \begin{pmatrix} m_{11}(X) & -m_{12}(X) & m_{13}(X) \\
-m_{21}(X) & m_{22}(X) & -m_{23}(X) \\
m_{31}(X) & -m_{32}(X) & m_{33}(X)
\end{pmatrix},$$
where $m_{ij}(X)$ denotes the $(ij)$th minor $X$, i.e. $m_{ij}(X)$ equals the determinant of the $2 \times 2$-matrix obtained from $X$ by deleting the $i$th row and the $j$th column.
 
It follows from (\ref{mulax}) that the eigenfunction $\mu^A$ satisfies the Lax pair
\begin{equation}\label{mmulax}
\begin{cases}
\mu^A_x + [\mathcal{L}, \mu^A] = -V_1^T\mu^A, \\
\mu^A_t + [\mathcal{Z}, \mu^A] = -V_2^T\mu^A.
\end{cases}
\end{equation}
Thus, the eigenfunctions $\{\mu_j^A\}_1^3$ are solutions of the integral equations
\begin{equation}\label{mujAdef}
  \mu_j^A(x,t,k) = I -  \int_{\gamma_j} e^{-\hat{\mathcal{L}} (x-x') - \hat{\mathcal{Z}}(t-t')} (V_1^T dx + V_2^T dt)\mu^A, \qquad j = 1, 2,3.
\end{equation}
The functions $\mu_1^A$ and $\mu_2^A$ are entire functions of $k$, whereas the first, second, and third columns of $\mu_3^A$, whose definitions involve integration from $x = \infty$, are defined for $k \in (\omega \hat{\mathcal{S}}, \omega^2 \hat{\mathcal{S}}, \hat{\mathcal{S}})$. Moreover, 
\begin{align*}
& \mu_2^A(x,t,k) \; \text{is bounded for} \; k \in  (\omega \hat{\mathcal{R}}, \omega^2 \hat{\mathcal{R}}, \hat{\mathcal{R}}),
	\\
& \mu_3^A(x,t,k) \; \text{is bounded for} \; k \in (\omega \hat{\mathcal{S}}, \omega^2 \hat{\mathcal{S}}, \hat{\mathcal{S}}),
\end{align*}
whereas the corresponding sets where the columns of $\mu_1^A$ are bounded are all empty.
It also holds that
\begin{align*}
& \mu_2^A(x, t, k) = I + O(1/k), \qquad k \to \infty, \quad k \in (\omega \hat{\mathcal{R}}, \omega^2 \hat{\mathcal{R}}, \hat{\mathcal{R}}),
	\\
& \mu_3^A(x,t,k) = I + O(1/k), \qquad k \to \infty, \quad k \in (\omega \hat{\mathcal{S}}, \omega^2 \hat{\mathcal{S}}, \hat{\mathcal{S}}).
\end{align*}	
For $x = 0$, $\mu_1^A$ and $\mu_2^A$ have the following enlarged domains of boundedness:
\begin{align*}
&\mu_1^A(0,t,k)  \; \text{is bounded for} \; k \in (\omega\mathcal{T} \cup \omega \hat{\mathcal{T}}, \omega^2\mathcal{T} \cup \omega^2 \hat{\mathcal{T}}, 
\mathcal{T} \cup \hat{\mathcal{T}}),
	\\
& \mu_2^A(0,t,k)  \; \text{is bounded for} \; k \in (\omega\mathcal{R} \cup \omega \hat{\mathcal{R}}, \omega^2\mathcal{R} \cup \omega^2 \hat{\mathcal{R}}, 
\mathcal{R} \cup \hat{\mathcal{R}}).
\end{align*}

\begin{remark}\upshape
1. The inverse of a unit determinant $3 \times 3$ matrix $X$ is given by $X^{-1} = (X^A)^T$.
The tracelessness of the matrices $L$ and $Z$ implies that $\det \mu_j(x,t,k) \equiv 1$, $k \in \C$, for $j = 1,2$.  

2. It is crucial that many of the minors $m_{ij}(\mu_k)$ have much larger domains of definitions and boundedness than the analysis in subsection \ref{mujsubsec} suggests. For example, we have seen that $[\mu_3]_2$ and $[\mu_3]_3$ are defined and bounded for $k$ in $\omega^2 \mathcal{S}$ and $\mathcal{S}$, respectively, where $[X]_j$ denotes the $j$th column of the matrix $X$. Nevertheless, the combination $m_{11}(\mu_3) = (\mu_3)_{22}(\mu_3)_{33} - (\mu_3)_{23}(\mu_3)_{32}$ is bounded not only for $k \geq 0$ but for all $k \in \omega \hat{\mathcal{S}}$.
\end{remark}

\subsection{Symmetries}
The $\Z_3$ symmetry (\ref{Z3symm}) implies corresponding symmetries for the eigenfunctions.

\begin{lemma}\label{symmlemma} 
The functions $M$ and $\{\mu_j\}_1^3$ obey the symmetries
\begin{align}\label{MZ3symm}
&  M(x,t,k) =  \mathcal{A} M(x,t,\omega k)\mathcal{A}^{-1}, \qquad k \in \C,
\end{align}
and
\begin{align}\nonumber
&  \mu_j(x,t,k) =  \mathcal{A} \mu_j(x,t,\omega k)\mathcal{A}^{-1}, &&  k \in \C \ \ \text{\upshape if} \ \ j = 1,2; 
	\\ \label{mujZ3symm}
& && k \in (\omega \mathcal{S}, \omega^2 \mathcal{S},\mathcal{S}) \ \ \text{\upshape if} \ \ j = 3.
\end{align}
\end{lemma}
\proofbegin
We will prove (\ref{MZ3symm}) for $k \in D_1$; the proofs when $k \in D_j$ for $j \neq 1$ are analogous. Thus, suppose that $k \in D_1$. Then $\omega k \in D_5$ and the functions 
$$\psi(x,t,k) = M_1(x,t,k)e^{\mathcal{L}(k) x + \mathcal{Z}(k)t} \quad \text{and} \quad 
\phi(x,t,k) = M_5(x,t,\omega k)e^{\mathcal{L}(\omega k) x + \mathcal{Z}(\omega k)t}$$ 
satisfy the equations
$$\begin{cases}
\psi_x(x,t,k) = L(x,t,k)\psi(x,t,k), \\
\psi_t(x,t,k) = Z(x,t,k) \psi(x,t,k),
\end{cases}
\qquad
\begin{cases}
\phi_x(x,t,k) = L(x,t,\omega k) \phi(x,t,k), \\
\phi_t(x,t,k) = Z(x,t,\omega k) \phi(x,t,k).
\end{cases}$$
In view of the symmetries (\ref{Z3symm}), this implies that $\psi(x,t,k)$ and $\mathcal{A} \phi(x,t,k) \mathcal{A}^{-1}$ satisfy the same ODE. Thus, the matrix $J(k)$ defined by
\begin{align}\nonumber
J(k) & = e^{-\mathcal{L}(k) x - \mathcal{Z}(k)t}M_1(x,t,k)^{-1}\mathcal{A} M_5(x,t,\omega k)e^{\mathcal{L}(\omega k)x + \mathcal{Z}(\omega k)t}\mathcal{A}^{-1}
	\\ \label{M1M5}
& = e^{-\hat{\mathcal{L}}(k) x - \hat{\mathcal{Z}}(k)t}\left(M_1(x,t,k)^{-1}\mathcal{A} M_5(x,t,\omega k)\mathcal{A}^{-1}\right), \qquad k \in D_1,
\end{align}
is independent of $x$ and $t$. We have to show that $J(k) = I$.
By (\ref{gammaijnudef}), the matrices $\gamma^1$ and $\gamma^5$ defined by $(\gamma^1)_{ij} := \gamma_{ij}^{1}$ and $(\gamma^5)_{ij} := \gamma_{ij}^{5}$ are given by
\begin{align}\label{gamma1gamma5}
\gamma^1 = \begin{pmatrix} \gamma_3 & \gamma_3 & \gamma_3 \\
\gamma_1 & \gamma_3 & \gamma_3 \\
\gamma_1 & \gamma_2 & \gamma_3 \end{pmatrix}, \qquad
\gamma^5 = \begin{pmatrix} \gamma_3 & \gamma_1 & \gamma_2 \\
\gamma_3 & \gamma_3 & \gamma_3 \\
\gamma_3 & \gamma_1 & \gamma_3 \end{pmatrix}.
\end{align}
Consequently, 
\begin{align}\label{M1at00}
M_1(0,T,k)
= 
\begin{pmatrix} * & * & * \\
0 & * & * \\
0 & * & * 
\end{pmatrix}, \qquad 
M_5(0,T,\omega k)
= 
\begin{pmatrix} * & 0 & * \\
* & * & * \\
* & 0 & * 
\end{pmatrix}, \qquad k \in D_1,
\end{align}
where $*$ denotes an unspecified entry. Evaluating (\ref{M1M5}) at $(x,t) = (0,T)$ and using (\ref{M1at00}) as well as the determinant condition $\det M_1 = 1$, we find that $J(k)$ has the form
$$J(k) =  \begin{pmatrix} * & * & * \\
0 & * & * \\
0 & * & * 
\end{pmatrix}.$$
Similarly, evaluating (\ref{M1M5}) as $x \to\infty$ and using that
$$\lim_{x \to \infty} M_1(x,0,k)
= 
\begin{pmatrix} 1 & 0 & 0 \\
* & 1 & 0 \\
* & * & 1 
\end{pmatrix}, \qquad 
\lim_{x \to \infty} M_5(x,0,\omega k)
= 
\begin{pmatrix} 1 & * & * \\
0 & 1 & 0 \\
0 & * & 1 
\end{pmatrix}, \qquad k \in D_1,
$$
we find that $J(k)$ has the form
$$J(k) =  \begin{pmatrix} 1 & 0 & 0 \\
* & 1 & 0 \\
* & * & 0
\end{pmatrix}.$$
It therefore only remains to show that $J_{32}(k) = 0$. Equation (\ref{gamma1gamma5}) yields
\begin{align}\label{M132M513}
(M_1(0,0,k))_{32} = 0, \qquad (M_5(0,0,\omega k))_{13} = 0.
\end{align} 
Evaluating the relation 
$$M_1(x,t,k)e^{\hat{\mathcal{L}}(k) x + \hat{\mathcal{Z}}(k)t}\begin{pmatrix} 1 & 0 & 0 \\ 0 & 1 & 0 \\ 0 & J_{32}(k) & 1Ê\end{pmatrix} = \mathcal{A} M_5(x,t,\omega k)\mathcal{A}^{-1}, \qquad k \in D_1,$$
at $(x,t) = (0,0)$ and using (\ref{M132M513}), we infer from the $(32)$-entry that
$$J_{32}(k) (M_1(0,0,k))_{33} = 0, \qquad k \in D_1.$$
Since $J_{32}(k)$ is an analytic function of $k \in D_1$ and $ (M_1(0,0,k))_{33} \to 1$ as $k \to \infty$ in $D_1$, we conclude that $J_{32}$ vanishes identically in $D_1$.
This proves (\ref{MZ3symm}) for $k \in D_1$.

Equation (\ref{mujZ3symm}) follows from the initial conditions
$$\mu_j(x_j, t_j, k) = I, \qquad j = 1,2,3,$$
and the symmetry properties (\ref{Z3symm}) of $L$ and $Z$ in a similar, but simpler, way.
\proofend

The symmetry (\ref{MZ3symm}) implies in particular that
\begin{align}\label{Snsymm}
  S_n(k) = \mathcal{A}S_{n+4}(\omega k)\mathcal{A}^{-1} , \qquad k \in D_n, \; n = 1, \dots, 12, 
\end{align}
and
\begin{align}\label{Jmnsymm}
  J_{n+1,n}(x,t,k) = \mathcal{A}J_{n+5,n+4}(x,t,\omega k))\mathcal{A}^{-1}, \qquad k \in D_n \cap D_{n+1}, \; n = 1, \dots, 12, 
\end{align}
where the index $n$ is defined modulo $12$.

\subsection{A matrix factorization problem}
Define the $3 \times 3$-matrix valued spectral functions $s(k)$ and $S(k)$ by
\begin{align}\label{mu3mu2mu1sS}
& \mu_3(x,t,k) = \mu_2(x,t,k) e^{\hat{\mathcal{L}}(k)x + \hat{\mathcal{Z}}(k)t} s(k), \qquad k \in (\omega \mathcal{S}, \omega^2 \mathcal{S}, \mathcal{S}),
	\\ \nonumber
& \mu_1(x,t,k) = \mu_2(x,t,k) e^{\hat{\mathcal{L}}(k)x + \hat{\mathcal{Z}}(k)t} S(k), \qquad k \in \C,
\end{align}
i.e.
\begin{align}\label{sSdef}  
  s(k) = \mu_3(0,0,k), \qquad S(k) = \mu_1(0,0,k).
\end{align}
We deduce from the properties of the $\mu_j$'s that $s(k)$ and $S(k)$ have the following boundedness properties:
\begin{align*}
& s(k)\; \text{is bounded and analytic for} \; k \in (\omega \mathcal{S}, \omega^2 \mathcal{S}, \mathcal{S}),
	\\
&S(k)\; \text{is bounded and analytic for} \; k \in (\omega\mathcal{R} \cup \omega \hat{\mathcal{R}}, \omega^2\mathcal{R} \cup \omega^2 \hat{\mathcal{R}}, 
\mathcal{R} \cup \hat{\mathcal{R}}),
	\\
&s^A(k)\; \text{is bounded and analytic for} \; k \in (\omega \hat{\mathcal{S}}, \omega^2 \hat{\mathcal{S}}, \hat{\mathcal{S}}),
	\\
&S^A(k) \; \text{is bounded and analytic for} \; k \in (\omega\mathcal{T} \cup \omega \hat{\mathcal{T}}, \omega^2\mathcal{T} \cup \omega^2 \hat{\mathcal{T}}, 
\mathcal{T} \cup \hat{\mathcal{T}}).
\end{align*}
Moreover,
\begin{align}\label{Mnmu2relation}
M_n(x,t,k) = \mu_2(x,t,k) e^{\hat{\mathcal{L}}(k)x + \hat{\mathcal{Z}}(k)t} S_n(k), \qquad k \in D_n.
\end{align}

\begin{proposition}\label{Snexplicitlemma}
 The $S_n$'s defined in (\ref{Sndef}) can be expressed in terms of the entries of $s(k)$ and $S(k)$ as follows:
 \begin{subequations}\label{Snexplicit}
\begin{align}\label{S1S2explicit}
&  S_1(k) = \begin{pmatrix}
\frac{S_{11}}{W_1} &  \frac{m_{21}(s)}{s_{33}} & s_{13} \\
 \frac{S_{21}}{W_1} &  \frac{m_{11}(s)}{s_{33}} & s_{23} \\
 \frac{S_{31}}{W_1} & 0 & s_{33}
  \end{pmatrix},
\qquad
  S_2(k) =  \begin{pmatrix}
 \frac{m_{22}(S)}{W_2} & \frac{m_{21}(s)}{s_{33}} & s_{13} \\
 \frac{m_{12}(S)}{W_2} & \frac{m_{11}(s)}{s_{33}} & s_{23} \\
 0 & 0 & s_{33}
\end{pmatrix},
  	\\ \label{S3S4explicit}
&  S_3(k) =  \begin{pmatrix}
 \frac{m_{22}(s)}{s_{33}} & \frac{m_{21}(S)}{W_3} & s_{13} \\
 \frac{m_{12}(s)}{s_{33}} & \frac{m_{11}(S)}{W_3} & s_{23} \\
 0 & 0 & s_{33}
   \end{pmatrix},
\qquad
  S_4(k) =  \begin{pmatrix}
 \frac{m_{22}(s)}{s_{33}} & \frac{S_{12}}{W_4} & s_{13} \\
 \frac{m_{12}(s)}{s_{33}} & \frac{S_{22}}{W_4} & s_{23} \\
 0 & \frac{S_{32}}{W_4} & s_{33}
   \end{pmatrix},
\end{align}
\end{subequations}
where the functions $\{W_j(k)\}_1^4$ are defined by
\begin{align*}
& W_1(k) = S_{11}m_{11}(s) - S_{21}m_{21}(s) + S_{31}m_{31}(s) = (S^Ts^A)_{11},
	\\
& W_2(k) = m_{11}(s)m_{22}(S) - m_{21}(s)m_{12}(S),
	\\
& W_3(k) = m_{22}(s)m_{11}(S) - m_{12}(s)m_{21}(S),
	\\
& W_4(k) = -S_{12}m_{12}(s) + S_{22}m_{22}(s) - S_{32}m_{32}(s) = (S^Ts^A)_{22}.
\end{align*}
The functions $\{S_n(k)\}_5^{12}$ can be obtained from (\ref{Snexplicit}) and the symmetries in (\ref{Snsymm}). 
\end{proposition}
\proofbegin
It will be convenient to work with eigenfunctions which are defined for all values of $k \in \C$. Thus, for each $X_0 > 0$, let $\gamma_3^{X_0}$ denote the contour $(X_0, 0) \to (x,t)$ in the $(x,t)$-plane. We introduce $\mu_3(x,t,k; X_0)$ as the solution of (\ref{mujdef}) with $j = 3$ and with the contour $\gamma_3$ replaced by $\gamma_3^{X_0}$. Similarly, we define $M_n(x,t,k; X_0)$, $n = 1, \dots, 12$, as the solutions of (\ref{Mnintegraleq}) with $\gamma_3$ replaced by $\gamma_3^{X_0}$. We will first derive expressions for $S_n(k; X_0) := M_n(0,0,k; X_0)$ in terms of $S(k)$ and $s(k; X_0) := \mu_3(0,0,k; X_0)$. Then (\ref{Snexplicit}) will follow by taking the limit $X_0 \to \infty$.

Define $R_n(k;X_0)$ and $T_n(k;X_0)$, $n = 1, \dots, 12$, by
\begin{subequations}\label{Sn13def}  
\begin{align}
 & R_n(k;X_0) = e^{- \hat{\mathcal{Z}}T}M_n(0, T,k;X_0), \qquad k \in D_n, 
 	\\
  &T_n(k; X_0) = e^{-\hat{\mathcal{L}} X_0}M_n(X_0, 0,k;X_0), \qquad k \in D_n.
\end{align}
\end{subequations}
Then 
\begin{align}\label{MnmujX0}
  \begin{cases}  M_n(x,t,k;X_0) = \mu_1(x,t,k) e^{\hat{\mathcal{L}}x + \hat{\mathcal{Z}}t} R_n(k;X_0), \\
  M_n(x,t,k;X_0) = \mu_2(x,t,k) e^{\hat{\mathcal{L}}x + \hat{\mathcal{Z}}t} S_n(k; X_0), \\
  M_n(x,t,k;X_0) = \mu_3(x,t,k;X_0) e^{\hat{\mathcal{L}}x + \hat{\mathcal{Z}}t} T_n(k;X_0),
  \end{cases} \qquad n = 1, \dots, 12, \; k \in D_n.
\end{align}  
The relations (\ref{MnmujX0}) imply that 
\begin{equation}\label{sSSnrelations}  
  s(k; X_0) = S_n(k; X_0)T_n^{-1}(k; X_0), \qquad S(k) = S_n(k; X_0)R_n^{-1}(k; X_0), \qquad k \in D_n.
\end{equation}
These equations constitute a matrix factorization problem which, given $s$ and $S$, can be solved for $\{R_n, S_n, T_n\}$. Indeed, the integral equations (\ref{Mnintegraleq}) together with the definitions of $\{R_n, S_n, T_n\}$ imply that
\begin{align} \nonumber
& \left(R_n(k; X_0)\right)_{ij} = 0 \quad \text{if} \quad \gamma_{ij}^n = \gamma_1,
	\\\label{Sn123ij0}
& \left(S_n(k; X_0)\right)_{ij} = 0 \quad \text{if} \quad \gamma_{ij}^n = \gamma_2,
	\\ \nonumber
& \left(T_n(k; X_0)\right)_{ij} = \delta_{ij} \quad \text{if} \quad \gamma_{ij}^n = \gamma_3,
\end{align}
where $\gamma_{ij}^n$ is defined by (\ref{gammaijnudef}). It follows that (\ref{sSSnrelations}) are $18$ scalar equations for $18$ unknowns. By computing the explicit solution of this algebraic system, we find that $\{S_n(k; X_0)\}_1^4$ are given by the equation obtained from (\ref{Snexplicit}) by replacing $\{S_n(k), s(k)\}$ with $\{S_n(k;  X_0), s(k; X_0)\}$. Taking $X_0 \to \infty$ in this equation,\footnote{Note that all quantities in this equation have well-defined limits as $X_0 \to \infty$. Had we instead tried to let $X_0 \to \infty$ already in (\ref{sSSnrelations}), the limit of the first equation in (\ref{sSSnrelations}) would only have been well-defined for $k \in (\omega \mathcal{S}, \omega^2 \mathcal{S}, \mathcal{S})$. This is the reason for introducing $X_0$.}
 we arrive at (\ref{Snexplicit}).
\proofend

\subsection{The global relation}
The spectral functions $S(k)$ and $s(k)$ are not independent but satisfy an important relation. 
Indeed, it follows from (\ref{mu3mu2mu1sS}) that
$$\mu_1 e^{\hat{\mathcal{L}}x + \hat{\mathcal{Z}}t} S^{-1}s = \mu_3.$$
Since $\mu_1(0,T,k) = I$, evaluation at $(0,T)$ yields the following {\it global relation}:
\begin{equation}\label{GR}
  S^{-1}(k)s(k) =  e^{-\hat{\mathcal{Z}}(k) T}c(T, k), \qquad k \in (\omega \mathcal{S}, \omega^2 \mathcal{S},\mathcal{S}),
\end{equation}
where the first, second, and thirds column vectors of the $3\times 3$-matrix valued function $c(T, k) := \mu_3(0,T,k)$ are analytic and of order $O(1/k)$ as $k \to \infty$ for $k$ in $\omega \mathcal{S}$, $\omega^2 \mathcal{S}$, and $\mathcal{S}$, respectively.
\begin{remark}\upshape
We also have
$$\mu_3^A = \mu_2^A e^{-\hat{\mathcal{L}}x - \hat{\mathcal{Z}}t} s^A, \qquad
\mu_1^A = \mu_2^A e^{-\hat{\mathcal{L}}x - \hat{\mathcal{Z}}t} S^A,$$
and so
$$\mu_1^Ae^{-\hat{\mathcal{L}}x - \hat{\mathcal{Z}}t}(S^{A-1}s^A) = \mu_3^A.$$
Evaluating this equation at $(0,T)$ we find the following cofactor version of the global relation:
\begin{align}\label{adGR}
S^T(k) s^A(k) = e^{\hat{\mathcal{Z}}(k)T}c^A(T, k), \qquad k \in (\omega \hat{\mathcal{S}}, \omega^2 \hat{\mathcal{S}}, \hat{\mathcal{S}}),
\end{align}
where the first, second, and thirds column vectors of the $3\times 3$-matrix valued function $c^A(T, k) := \mu_3^A(0,T,k)$ are analytic and of order $O(1/k)$ as $k \to \infty$ for $k$ in $\omega \hat{\mathcal{S}}$, $\omega^2 \hat{\mathcal{S}}$, and $\hat{\mathcal{S}}$, respectively.
\end{remark}

\subsection{The residue conditions}
Since $\mu_2$ is an entire function, it follows from (\ref{Mnmu2relation}) that $M$ can only have singularities at the points where the $S_n$'s have singularities. 
In view of the symmetries of Lemma \ref{symmlemma}, it is enough to study the singularities for $k \in \mathcal{S}$. We infer from the explicit formulas (\ref{Snexplicit}) that the possible singularities of $M$ in $\mathcal{S}$ are as follows:
\begin{itemize}
\item $[M]_2$ could have poles in $D_1 \cup D_2$ at the zeros of $s_{33}(k)$.
\item $[M]_1$ could have poles in $D_1$ at the zeros of $W_1(k)$.
\item $[M]_1$ could have poles in $D_2$ at the zeros of $W_2(k)$.
\item $[M]_1$ could have poles in $D_3 \cup D_4$ at the zeros of $s_{33}(k)$.
\item $[M]_2$ could have poles in $D_3$ at the zeros of $W_3(k)$.
\item $[M]_2$ could have poles in $D_4$ at the zeros of $W_4(k)$.
\end{itemize}
We denote the above possible zeros by $\{k_j\}_1^N$ and assume they satisfy the following assumption.

\begin{assumption}\label{kjass}\upshape
We assume that
\begin{itemize}
\item $s_{33}(k)$ has $n_1 \geq 0$ simple zeros in $D_1 \cup D_2$ denoted by $\{k_j\}_1^{n_1}$,
\item $W_1(k)$ has $n_2 - n_1\geq 0$ simple zeros in $D_1$ denoted by $\{k_j\}_{n_1 + 1}^{n_2}$,
\item $W_2(k)$ has $n_3 - n_2\geq 0$ simple zeros in $D_2$ denoted by $\{k_j\}_{n_2 + 1}^{n_3}$,
\item $s_{33}(k)$ has $n_4 - n_3\geq 0$ simple zeros in $D_3 \cup D_4$ denoted by $\{k_j\}_{n_3 + 1}^{n_4}$,
\item $W_3(k)$ has $n_5 - n_4\geq 0$ simple zeros in $D_3$ denoted by $\{k_j\}_{n_4 + 1}^{n_5}$,
\item $W_4(k)$ has $N - n_5\geq 0$ simple zeros in $D_4$ denoted by $\{k_j\}_{n_5 + 1}^{N}$,
\end{itemize}
and that none of these zeros coincide. Moreover, we assume that none of these functions have zeros on the boundaries of the $D_n$'s.
\end{assumption}

In the next proposition we determine the residue conditions at these zeros.

\begin{proposition}
Let $\{M_n\}_1^{12}$ be the eigenfunctions defined by (\ref{Mnintegraleq}) and assume that the set $\{k_j\}_1^N$ of singularities in $\mathcal{S}$ are as in assumption \ref{kjass}. Then the following residue conditions hold: 
\begin{subequations}\label{Mres}
\begin{align} \label{Mresa}
\underset{k_j}{\res} [M]_2 = &\; \frac{m_{21}(s(k_j))}{s_{13}(k_j)\dot{s}_{33}(k_j)} e^{\theta_{32}(k_j)} [M(k_j)]_3, \qquad 1 \leq j \leq n_1, \;k_j \in D_1 \cup D_2,
	\\ \nonumber
\underset{k_j}{\res} [M]_1 
  = &\;\frac{1}{\dot{W}_1(k_j)}\biggl\{ \frac{S_{11}(k_j)s_{23}(k_j) - S_{21}(k_j)s_{13}(k_j)}{m_{31}(s(k_j))}e^{\theta_{21}(k_j)}[M(k_j)]_2
  	\\ \label{Mresb}
&  + \frac{S_{31}(k_j)}{s_{33}(k_j)} e^{\theta_{31}(k_j)}[M(k_j)]_3  \biggr\}, \qquad n_1 < j \leq n_2, \; k_j \in D_1,
	\\ \label{Mresc}
\underset{k_j}{\res} [M]_1 
  = &\; \frac{m_{22}(S(k_j))s_{33}(k_j)}{\dot{W}_2(k_j)m_{21}(s(k_j))} e^{\theta_{21}(k_j)}[M(k_j)]_2, \qquad n_2 < j \leq n_3, \; k_j \in D_2,
	\\ \label{Mresd}
\underset{k_j}{\res} [M]_1 = &\; \frac{m_{22}(s(k_j))}{s_{13}(k_j)\dot{s}_{33}(k_j)} e^{\theta_{31}(k_j)} [M(k_j)]_3, \qquad n_3 < j \leq n_4,\;k_j \in D_3 \cup D_4,
	\\ \label{Mrese}
\underset{k_j}{\res} [M]_2 = &\; 
\frac{m_{21}(S(k_j))s_{33}(k_j)}{\dot{W}_3(k_j)m_{22}(s(k_j))}e^{\theta_{12}(k_j)}[M(k_j)]_1,
 \qquad n_4  < j \leq n_5, \; k_j \in D_3,
	\\ \nonumber
\underset{k_j}{\res} [M]_2 = &\; \frac{1}{\dot{W}_4(k_j)}\biggl\{\frac{S_{12}(k_j)s_{23}(k_j) - S_{22}(k_j)s_{13}(k_j)}{ m_{32}(s(k_j))}e^{\theta_{12}(k_j)} [M(k_j)]_1
	\\  \label{Mresf}
 &+ \frac{S_{32}(k_j)}{s_{33}(k_j)} e^{\theta_{32}(k_j)} [M(k_j)]_3 \biggr\},
 \qquad n_5 < j \leq N, \; k_j \in D_4,
\end{align}
\end{subequations}
where $\dot{f} := df/dk$, $\theta_{ij}$ is defined by
\begin{align}\label{thetadef}  
  \theta_{ij}(x,t,k) = (l_i(k) - l_j(k))x + (z_i(k) - z_j(k))t, \qquad i,j= 1,2,3,
\end{align}
and we have suppressed the $(x,t)$-dependence for simplicity.
\end{proposition}
\proofbegin
We will prove (\ref{Mresa})-(\ref{Mresc}); the other conditions follow by similar arguments.
Equation (\ref{Mnmu2relation}) implies the relations
\begin{equation}\label{M1mu2rel}  
  M_1 = \mu_2 e^{\hat{\mathcal{L}} x + \hat{\mathcal{Z}} t} S_1 \quad \text{and} \quad
  M_2 = \mu_2 e^{\hat{\mathcal{L}} x + \hat{\mathcal{Z}} t} S_2.
\end{equation}
In view of the expressions for $S_1$ and $S_2$ given in (\ref{S1S2explicit}), the three columns of (\ref{M1mu2rel}a) read
\begin{subequations}
\begin{align}\label{M1mu2a}
& [M_1]_1 = \frac{S_{11}}{W_1} [\mu_2]_1 
+ \frac{S_{21}}{W_1}e^{\theta_{21}} [\mu_2]_2
+ \frac{S_{31}}{W_1}e^{\theta_{31}} [\mu_2]_3,
	\\ \label{M1mu2b}
 & [M_1]_2 = \frac{m_{21}(s)}{s_{33}} e^{\theta_{12}} [\mu_2]_1 
  + \frac{m_{11}(s)}{s_{33}} [\mu_2]_2,  
   	\\ \label{M1mu2c}
& [M_1]_3 = s_{13} e^{\theta_{13}} [\mu_2]_1 
 + s_{23} e^{\theta_{23}} [\mu_2]_2
  + s_{33}[\mu_2]_3,
\end{align}
\end{subequations} 
while the three columns of (\ref{M1mu2rel}b) read
 \begin{subequations}
\begin{align}\label{M2mu2a}
& [M_2]_1 = \frac{m_{22}(S)}{W_2} [\mu_2]_1 
+ \frac{m_{12}(S)}{W_2}e^{\theta_{21}} [\mu_2]_2
	\\ \label{M2mu2b}
 & [M_2]_2 = \frac{m_{21}(s)}{s_{33}} e^{\theta_{12}} [\mu_2]_1 
  + \frac{m_{11}(s)}{s_{33}} [\mu_2]_2,  
   	\\ \label{M2mu2c}
& [M_2]_3 = s_{13} e^{\theta_{13}} [\mu_2]_1 
 + s_{23} e^{\theta_{23}} [\mu_2]_2
  + s_{33}[\mu_2]_3.
\end{align}
\end{subequations}  

In order to prove (\ref{Mresa}), we first suppose thatÊ $k_j \in D_1$ is a simple zero of $s_{33}(k)$.
Solving (\ref{M1mu2c}) for $[\mu_2]_1$ and substituting the result into (\ref{M1mu2b}), we find
$$ [M_1]_2 = \frac{m_{21}(s)}{s_{13}s_{33}}e^{\theta_{32}}[M_1]_3
- \frac{m_{21}(s)}{s_{13}}e^{\theta_{32}}[\mu_2]_3
- \frac{m_{31}(s)}{s_{13}}[\mu_2]_2.$$
Taking the residue of this equation at $k_j$, we find the condition (\ref{Mresa}) in the case when $k_j \in D_1$. 
Similarly, solving (\ref{M2mu2c}), substituting the result into (\ref{M2mu2b}), and taking the residue at $k_j$, we find that (\ref{Mresa}) also holds if $k_j$ is a simple zero of $s_{33}$ in $D_2$.

In order to prove (\ref{Mresb}), we suppose thatÊ $k_j \in D_1$ is a simple zero of $W_1(k)$.
Solving (\ref{M1mu2b}) and (\ref{M1mu2c}) for $ [\mu_2]_1$ and $[\mu_2]_2$ and substituting the result into (\ref{M1mu2a}), we find
\begin{align}
[M_1]_1 
  = &\;\frac{e^{\theta_{31}} }{m_{31}(s)}[\mu_2]_3
    + \frac{S_{11}s_{23} - S_{21}s_{13}}{W_1 m_{31}(s)} e^{\theta_{21}}[M_1]_2
  + \left(\frac{S_{31}}{W_1 s_{33}}
  - \frac{1}{s_{33} m_{31}(s)}\right) e^{\theta_{31}}[M_1]_3.
\end{align}
Taking the residue of this equation at $k_j$, we find (\ref{Mresb}).

In order to prove (\ref{Mresc}), we suppose thatÊ $k_j \in D_2$ is a simple zero of $W_2(k)$.
Solving (\ref{M2mu2b}) for $ [\mu_2]_1$ and substituting the result into (\ref{M2mu2a}), we find
\begin{align}
[M_2]_1 
  = &\;- \frac{e^{\theta_{21}} }{m_{21}(s)}[\mu_2]_2
    + \frac{m_{22}(S)s_{33}}{W_2 m_{21}(s)} e^{\theta_{21}}[M_2]_2
 \end{align}
Taking the residue of this equation at $k_j$, we find (\ref{Mresc}).

\proofend

\section{The Riemann-Hilbert problem}\nequation\label{RHsec}
The sectionally analytic function $M(x,t,k)$ defined in section \ref{specanalysissec} satisfies a Riemann-Hilbert problem which can be formulated in terms of the initial and boundary values of $q(x,t)$ and $r(x,t)$. By solving this RH problem, the solution of (\ref{qrsystem}) can be recovered for all values of $x, t$.

\begin{theorem}\label{RHtheorem}
Suppose that $\{q(x,t), r(x,t)\}$ is a solution of the system (\ref{qrsystem}) in the half-line domain $\{0 < x < \infty, 0 < t < T\}$ with sufficient smoothness and decay as $x \to \infty$.
Then $u(x,t)$ can be reconstructed from the initial and boundary values $\{q_0(x), r_0(x), g_0(t), h_0(t), g_1(t), h_1(t)\}$ defined in (\ref{boundaryvalues}) as follows.

Use the initial and boundary data to define the jump matrices $J_{m,n}(x, t, k)$, $n,m = 1, \dots, 12$, by equation (\ref{Jmndef}) as well as the spectral functions $s(k)$ and $S(k)$ by equation (\ref{sSdef}). 
Assume that the possible zeros $\{k_j\}_1^N$ of the functions $s_{33}(k)$ and $\{W_j(k)\}_1^4$ in $\mathcal{S}$ are as in assumption \ref{kjass}. 
 
Then the solution $\{q(x,t), r(x,t)\}$ is given by
\begin{equation}\label{recoverqr}
q(x,t) = i\sqrt{3} \lim_{kÊ\to \infty} (kM(x,t,k))_{23}, \qquad
r(x,t) = -i \omega^2 \sqrt{3} \lim_{kÊ\to \infty} (kM(x,t,k))_{13}, 
\end{equation}
where $M(x,t,k)$ satisfies the following $3\times 3$ matrix RH problem:
\begin{itemize}
\item $M$ is sectionally meromorphic on the Riemann $k$-sphere with jumps across the contours $\bar{D}_n \cap \bar{D}_m$, $n, m = 1, \dots, 12$, see Figure \ref{Dns.pdf}.

\item Across the contours $\bar{D}_n \cap \bar{D}_m$, $n, m = 1, \dots, 12$, $M$ satisfies the jump condition (\ref{MnMmrelation}).

\item $M(x,t,k) = I + O\left(\frac{1}{k}\right), \qquad k \to \infty.$

\item The first column of $M$ has simple poles at $k = k_j$ for $n_1 < j \leq n_4$. The second column of $M$ has simple poles at $k = k_j$ for $1\leq  j \leq n_1$ and $n_4 < j \leq N$. The associated residues satisfy the relations in (\ref{Mres}).

\item For each zero $k_j$ in $\mathcal{S}$, there are two additional points,
$$\omega k_j, \quad \omega^2 k_j,$$
at which $M$ also has simple poles. The associated residues satisfy the residue conditions obtained from (\ref{Mres}) and the symmetries (\ref{MZ3symm}). 
\end{itemize}
\end{theorem}
\proofbegin
It only remains to prove (\ref{recoverqr}) and this equation follows from the large $k$ asymptotics of the eigenfunctions, see subsection \ref{asymptoticssubsec} below for details. 
\proofend

\section{Linearizable boundary conditions}\nequation\label{linearizablesec}
Theorem \ref{RHtheorem} expresses the solution $\{q(x,t), r(x,t)\}$ of the system (\ref{qrsystem}) in terms of the solution of a RH problem whose jump matrix is given in terms of the spectral functions $s(k)$ andÊ $S(k)$. The function $s(k)$ is defined in terms of the initial data $\{q_0(x), r_0(x)\}$ via a system of linear Volterra integral equations, whereas the function $S(k)$ is defined in terms of the boundary data $\{g_0(t), h_0(t), g_1(t), h_1(t)\}$ also via a system of linear Volterra integral equations. However, for a well-posed problem, only part of the boundary data can be independently prescribed; for example, for the Dirichlet problem $g_0$ and Ê$h_0$ are prescribed, whereas for the Neumann problem $g_1$ and $h_1$ are prescribed. 

In the next section, we will show that in general $S(k)$ can be determined from the given boundary conditions and the initial data via a system of {\it nonlinear} integral equations. In this section, we will consider the special case of {\it linearizable} boundary conditions.
These are boundary conditions for which it is possible to express $S(k)$ in terms of $s(k)$ and the given boundary conditions via algebraic manipulations. Thus, the above nonlinear step can be avoided and the problem can be fully solved by linear operations alone.

Our present goal is to find an approach for analyzing linearizable boundary conditions for equations with $3 \times 3$ Lax pairs. To this end, we will consider IBV problems for the system (\ref{qrsystem}) with arbitrary initial data and vanishing Dirichlet boundary conditions:
\begin{align}\label{linearizableBCs}
  q(0,t) = r(0,t) = 0, \qquad 0 < t < \infty.
\end{align}
We will show that these particular boundary conditions are linearizable in the sense that the jump matrix as well as the residue conditions of the RH problem of Theorem \ref{RHtheorem} can be expressed entirely in terms of $s(k)$. The elimination of $S(k)$ is achieved by analyzing the global relation and the equations obtained from the global relation under those transformations in the complex $k$-plane which leave the associated dispersion relation invariant. For the system (\ref{qrsystem}), there is one such transformation, namely $k \mapsto -k$ (which leaves $k^2$ invariant).

For a function $f(k)$, we define $\hat{f}(k)$ by $\hat{f}(k) = f(-k)$.

\begin{theorem}\label{linearizableth}
Let $\{q(x,t), r(x,t)\}$ satisfy the system (\ref{qrsystem}) in the half-line domain $\{0 < x < \infty,\; 0<t<\infty \}$ together with the initial conditions
$$q(x,0) = q_0(x), \quad r(x,0) = r_0(x), \qquad 0 < x < \infty,$$
and the Dirichlet boundary conditions
\begin{align}\label{vanishingDirichlet}
q(0,t) = r(0,t) = 0, \qquad 0 < t < \infty.
\end{align}
Assume that the initial and boundary conditions are compatible at $(x,t) = (0,0)$. 

Then the jump matrices $J_{m,n}(k)$ for the RH problem in Theorem \ref{RHtheorem} are given explicitly in terms of the spectral function $s(k)$ as follows: 
\begin{align}\nonumber
& J_{1,2} = \begin{pmatrix} 1 & 0 & 0 \\
 0 & 1 & 0 \\
 -\frac{e^{\theta_{31}}
   \hat{s}_{31}}{s_{33}\mathcal{W}_1} & 0
   & 1
   \end{pmatrix}, \quad
 J_{2,3} = \begin{pmatrix}
 1 & \frac{e^{\theta_{12}} (\hat{s}^Ts^A)_{21}}{\mathcal{W}_4} & 0 \\
- \frac{e^{\theta_{21}} (\hat{s}^Ts^A)_{12}}{\mathcal{W}_1} 
& -\frac{(\hat{s}^Ts^A)_{21}(\hat{s}^Ts^A)_{12}}{\mathcal{W}_1 \mathcal{W}_4} & 0 \\
 0 & 0 & 1
\end{pmatrix},
	\\ \label{Jlinearizable}
& J_{3,4} =  \begin{pmatrix}
1 & 0 & 0 \\
 0 & 1 & 0 \\
 0 & \frac{e^{\theta_{32}} \hat{s}_{32}}{ s_{33}\mathcal{W}_4} & 1
\end{pmatrix},
\qquad J_{4,5} = \begin{pmatrix}
 1 & 0 & -\frac{e^{\theta_{13}}
   s_{13}}{s_{11}} \\
 0 & 1 & 0 \\
 \frac{e^{\theta_{31}} s_{31}}{s_{33}}  & 0 & \frac{m_{22}(s)}{s_{11} s_{33}}
\end{pmatrix},
\end{align}
where $\theta_{ij}(x,t,k)$ is given by (\ref{thetadef}) and the functions $\{\mathcal{W}_1(k), \mathcal{W}_4(k)\}$ are defined by
\begin{subequations}
\begin{align}\label{calW1def}
& \mathcal{W}_1(k) = \hat{s}_{11} m_{11}(s)-\hat{s}_{21} m_{21}(s)+\hat{s}_{31}
   m_{31}(s),
	\\ \label{calW4def}
& \mathcal{W}_4(k) = - \hat{s}_{12} m_{12}(s) + \hat{s}_{22} m_{22}(s) - \hat{s}_{32} m_{32}(s).
\end{align}
\end{subequations}
The expressions for the remaining jump matrices follow from the expressions in (\ref{Jlinearizable}) and the symmetry (\ref{Jmnsymm}).
Moreover, the residue conditions (\ref{Mres}) are given in terms of $s(k)$ by
\begin{subequations}\label{Mreslinearizable}
\begin{align} \label{Mreslinearizablea}
\underset{k_j}{\res} [M]_2 = &\; \frac{m_{21}(s(k_j))}{s_{13}(k_j)\dot{s}_{33}(k_j)} e^{\theta_{32}(k_j)} [M(k_j)]_3, \qquad 1 \leq j \leq n_1, \;k_j \in D_1 \cup D_2,
	\\ \nonumber
\underset{k_j}{\res} [M]_1 
  = &\;\frac{1}{\dot{\mathcal{W}}_1(k_j)}\biggl\{\frac{\hat{s}_{11}(k_j)s_{23}(k_j) - \hat{s}_{21}(k_j)s_{13}(k_j)}{m_{31}(s(k_j))}e^{\theta_{21}(k_j)}[M(k_j)]_2
  	\\ \label{Mreslinearizableb}
&  + \frac{\hat{s}_{31}(k_j)}{s_{33}(k_j)} e^{\theta_{31}(k_j)}[M(k_j)]_3  \biggr\}, \qquad n_1 < j \leq n_2, \; k_j \in D_1,
	\\\nonumber
\underset{k_j}{\res} [M]_1 
  = &\; \frac{\hat{s}_{11}(k_j)s_{33}(k_j) - \hat{s}_{31}(k_j)s_{13}(k_j)}{\dot{\mathcal{W}}_1(k_j)m_{21}(s(k_j))} e^{\theta_{21}(k_j)}[M(k_j)]_2, 
  	\\ \label{Mreslinearizablec}
& \hspace{6cm}	 n_2 < j \leq n_3, \; k_j \in D_2,
	\\ \label{Mreslinearizabled}
\underset{k_j}{\res} [M]_1 = &\; \frac{m_{22}(s(k_j))}{s_{13}(k_j)\dot{s}_{33}(k_j)} e^{\theta_{31}(k_j)} [M(k_j)]_3, \qquad n_3 < j \leq n_4,\;k_j \in D_3 \cup D_4,
	\\ \nonumber
\underset{k_j}{\res} [M]_2 = &\; \frac{\hat{s}_{12}(k_j)s_{33}(k_j) - \hat{s}_{32}(k_j)s_{13}(k_j)}{\dot{\mathcal{W}}_4(k_j) m_{22}(s(k_j))}e^{\theta_{12}(k_j)}[M(k_j)]_1,
 	\\\label{Mreslinearizablee}
& \hspace{6cm}  n_4  < j \leq n_5, \; k_j \in D_3,
	\\ \nonumber
\underset{k_j}{\res} [M]_2 = &\; \frac{1}{\dot{\mathcal{W}}_4(k_j)}\biggl\{\frac{\hat{s}_{12}(k_j)s_{23}(k_j) - \hat{s}_{22}(k_j)s_{13}(k_j)}{m_{32}(s(k_j))}e^{\theta_{12}(k_j)} [M(k_j)]_1
	\\  \label{Mreslinearizablef} 
&+ \frac{\hat{s}_{32}(k_j)}{s_{33}(k_j)} e^{\theta_{32}(k_j)} [M(k_j)]_3 \biggr\},
 \qquad n_5 < j \leq N, \; k_j \in D_4.
\end{align}
\end{subequations}
\end{theorem}
\proofbegin
The vanishing boundary conditions (\ref{vanishingDirichlet}) imply that the matrix $Z$ defined in (\ref{LZdef}) satisfies $Z(x,t,k) = Z(x,t,-k)$. It follows that the spectral function $S(k)$ obeys the symmetry
\begin{align}\label{SkSminusk}  
  S(k) = S(-k), \qquad k \in \C.
\end{align}
We will use this symmetry property together with the global relation (\ref{GR}) to eliminate all the entries of $S(k)$ from the jump matrices and the residue conditions. 

Since we are now assuming that $T= \infty$, the columns of $S(k)$ have smaller domains of definition than above (but the domains of boundedness remain the same). More precisely, 
\begin{align}
S(k) \;\text{is defined and bounded for} \; k \in (\omega\mathcal{R} \cup \omega \hat{\mathcal{R}}, \omega^2\mathcal{R} \cup \omega^2 \hat{\mathcal{R}}, 
\mathcal{R} \cup \hat{\mathcal{R}}).
\end{align}
The global relation when $T = \infty$ is
\begin{align}\label{GRinfty}
(S^A)^Ts = \begin{pmatrix}
c_{11} & 0 & 0 \\
0 & c_{22}  & 0 \\
0 & 0 & c_{33}  
\end{pmatrix}, \qquad k \in \begin{pmatrix}\emptyset & D_9\cup D_{10} & D_3 \cup D_{4} \\
D_7 \cup D_8 & \emptyset & D_1 \cup D_2 \\
D_{5} \cup D_{6} & D_{11} \cup D_{12} & \emptyset \end{pmatrix},
\end{align}
where 
$$(S^A)^T(k) = \begin{pmatrix} m_{11}(S) & -m_{21}(S) & m_{31}(S) \\
-m_{12}(S) & m_{22}(S) & -m_{32}(S) \\
m_{13}(S) & -m_{23}(S) & m_{33}(S)
\end{pmatrix},$$
and the notation indicates that the $(11)$ entry of (\ref{GRinfty}) is not valid for any $k \in \C$, the $(12)$ entry of (\ref{GRinfty}) is valid for $k \in D_9 \cup D_{10}$ etc.

We first derive the expression for $J_{2,3}$. 
From the explicit expressions in (\ref{Snexplicit}), we see that the matrices $S_2(k)$ and $S_3(k)$ only depend on the entries ofÊ $S(k)$ via the quotients
$$\frac{m_{12}(S)}{m_{22}(S)} \quad \text{and} \quad 
\frac{m_{11}(S)}{m_{21}(S)},$$
respectively.
Because of the symmetry (\ref{SkSminusk}), the global relation (\ref{GRinfty}) yields {\it two} algebraic relations for the entries of $S(k)$ in each of the $D_j$'s. In $D_2$, these relations are 
\begin{subequations}\label{D2relations}
\begin{align}\label{D2relation1}
-s_{13} m_{12}(S)+s_{23} m_{22}(S)-s_{33}
   m_{32}(S)=0,
	\\ \label{D2relation2}
-\hat{s}_{11} m_{12}(S)+\hat{s}_{21} m_{22}(S)-\hat{s}_{31}
   m_{32}(S)=0.
 \end{align}
\end{subequations}
Equation (\ref{D2relation1}) is the $(23)$ entry of (\ref{GRinfty}), whereas (\ref{D2relation2}) is obtained by letting $k \to -k$ in the $(21)$ entry of (\ref{GRinfty}) and using (\ref{SkSminusk}) in the resulting equation.
The two relations in (\ref{D2relations}) can be solved for the two quotients 
$$\frac{m_{12}(S)}{m_{32}(S)} \quad  \text{and} \quad\frac{m_{22}(S)}{m_{32}(S)}$$
with the result that
\begin{subequations}\label{D2quotients}
\begin{align}
& \frac{m_{12}(S)}{m_{32}(S)} = \frac{s_{23} \hat{s}_{31}-\hat{s}_{21} s_{33}}{s_{13}
   \hat{s}_{21}-\hat{s}_{11} s_{23}}, \qquad k \in D_2,
	\\
& \frac{m_{22}(S)}{m_{32}(S)} 
= \frac{s_{13} \hat{s}_{31}-\hat{s}_{11} s_{33}}{s_{13}
   \hat{s}_{21}-\hat{s}_{11} s_{23}}, \qquad k \in D_2.
\end{align}
\end{subequations}
Using these relations to eliminate the $S(k)$ dependence in the expression for $S_2$ in (\ref{S1S2explicit}), we find
\begin{align}\label{S2linearizable}
S_2(k) = \begin{pmatrix}
 \frac{\hat{s}_{11} s_{33}-s_{13} \hat{s}_{31}}{s_{33}
   (\hat{s}_{11} m_{11}(s)-\hat{s}_{21}
   m_{21}(s)+\hat{s}_{31} m_{31}(s))} &
   \frac{m_{21}(s)}{s_{33}} & s_{13} \\
 \frac{\hat{s}_{21} s_{33}-s_{23} \hat{s}_{31}}{s_{33}
   (\hat{s}_{11} m_{11}(s)-\hat{s}_{21}
   m_{21}(s)+\hat{s}_{31} m_{31}(s))} &
   \frac{m_{11}(s)}{s_{33}} & s_{23} \\
 0 & 0 & s_{33}
 \end{pmatrix}.
\end{align}

Similarly, the $(13)$ entry of (\ref{GRinfty}) together with the equation obtained by letting $k \to -k$ in the $(12)$ entry of (\ref{GRinfty}) yield two relations for the entries of $S(k)$ in $D_3$. Solving these two relations for the two quotients
$$\frac{m_{11}(S)}{m_{31}(S)} \quad \text{and} \quad
  \frac{m_{21}(S)}{m_{31}(S)},$$ 
we find
\begin{subequations}\label{D3quotients}
\begin{align}
& \frac{m_{11}(S)}{m_{31}(S)}
= \frac{s_{23} \hat{s}_{32}-\hat{s}_{22} s_{33}}{s_{13}
   \hat{s}_{22}-\hat{s}_{12} s_{23}}, \qquad k \in D_3,
	\\
& \frac{m_{21}(S)}{m_{31}(S)} = \frac{s_{13} \hat{s}_{32}-\hat{s}_{12} s_{33}}{s_{13}  \hat{s}_{22}-\hat{s}_{12} s_{23}}, \qquad k \in D_3.
\end{align}
\end{subequations}
Using these relations to eliminate the $S(k)$ dependence in the expression for $S_3$ in (\ref{S3S4explicit}), we find
\begin{align}\label{S3linearizable}
S_3(k) = \begin{pmatrix}
 \frac{m_{22}(s)}{s_{33}} & \frac{s_{13} \hat{s}_{32}-\hat{s}_{12} s_{33}}{s_{33}(\hat{s}_{12} 
   m_{12}(s)-\hat{s}_{22} m_{22}(s)+\hat{s}_{32}m_{32}(s))} & s_{13} \\
 \frac{m_{12}(s)}{s_{33}} & \frac{s_{23} \hat{s}_{32}-\hat{s}_{22} s_{33}}{s_{33}(\hat{s}_{12}m_{12}(s)-\hat{s}_{22} m_{22}(s)+\hat{s}_{32} m_{32}(s))} & s_{23} \\
 0 & 0 & s_{33}
  \end{pmatrix}.
\end{align}
The expression for $J_{2,3}$ in (\ref{Jlinearizable}) now follows by substituting the matrices (\ref{S2linearizable}) and (\ref{S3linearizable}) into the definition $J_{2,3} = e^{\hat{\mathcal{L}}x + \hat{\mathcal{Z}}t}(S_2^{-1}S_3)$, recalling that both matrices have unit determinant.

We next derive the expression for $J_{1,2}$. The approach used for finding $J_{2,3}$ is not successful in this case (see remark \ref{linearizableremark} below), and we have to rely on a somewhat different argument. 

The explicit expression for the matrix $S_2(k)$ in (\ref{S1S2explicit}) only involves entries from the following column vectors:
$$[s]_3, \quad [S^A]_2, \quad [s^A]_1.$$
Since all these column vectors are analytic in $D_1$, $S_2(k)$ admits an analytic continuation to $D_1$. It follows that the defining equation
\begin{align}\label{J12defined}
J_{1,2} = e^{\hat{\mathcal{L}}x + \hat{\mathcal{Z}}t}(S_1^{-1}S_2),
\end{align}
which a priori is valid only for $k \in \bar{D}_1 \cap \bar{D}_2$, can be extended to the line $k \in \bar{D}_{12} \cap \bar{D}_1$. 
Substituting the explicit expressions in (\ref{Snexplicit}) for $S_1(k)$ and $S_2(k)$ into (\ref{J12defined}), we find after some algebra that $J_{1,2}$ depends on the entries ofÊ $S(k)$ only via the four quotients
\begin{align}\label{J12quotients}
\frac{m_{13}(S)}{m_{33}(S)}, \quad 
\frac{m_{23}(S)}{m_{33}(S)},\quad
\frac{m_{12}(S)}{m_{32}(S)}, \quad
\frac{m_{22}(S)}{m_{32}(S)}.  
\end{align}

The $(32)$ entry of (\ref{GRinfty}) together with the equation obtained by letting $k \to -k$ in the $(31)$ entry of (\ref{GRinfty}) yield two relations for the entries of $S(k)$ in $D_{12}$. Solving these two relations for the first pair of quotients in (\ref{J12quotients}), we find
\begin{align}
& \frac{m_{13}(S)}{m_{33}(S)}
= \frac{s_{22} \hat{s}_{31}-\hat{s}_{21} s_{32}}{s_{12}
   \hat{s}_{21}-\hat{s}_{11} s_{22}}, \qquad k \in D_{12},
	\\
& \frac{m_{23}(S)}{m_{33}(S)} = \frac{s_{12} \hat{s}_{31}-\hat{s}_{11} s_{32}}{s_{12}
   \hat{s}_{21}-\hat{s}_{11} s_{22}}, \qquad k \in D_{12}.
\end{align}
We use these quotients to eliminate the first two quotients in (\ref{J12quotients}) from the expression for $J_{1,2}$ for $k \in \bar{D}_{12} \cap \bar{D}_1$.

On the other hand, since all quantities in the equations (\ref{D2quotients}) are analytic in $D_1$, the relations in (\ref{D2quotients}) are valid also for $k \in \bar{D}_{12} \cap \bar{D}_1$ and can be used to eliminate the second pair of quotients in (\ref{J12quotients}) from the expression for $J_{1,2}$ for $k \in \bar{D}_{12} \cap \bar{D}_1$.
This yields after simplification the following expression for $J_{1,2}$ for $k \in \bar{D}_{12} \cap \bar{D}_1$:
$$
J_{1,2} = \begin{pmatrix} 1 & 0 & 0 \\
 0 & 1 & 0 \\
 -\frac{e^{\theta_{31}}
   \hat{s}_{31}}{s_{33}(\hat{s}_{11} m_{11}(s)-\hat{s}_{21}
    m_{21}(s)+\hat{s}_{31}  m_{31}(s))} & 0
   & 1
   \end{pmatrix}, \qquad k \in \bar{D}_{12} \cap \bar{D}_1.$$
Since all quantities on the right-hand side are analytic in $D_1$, this expression for $J_{1,2}$ is valid also for $k \in \bar{D}_1 \cap \bar{D}_2$. This establishes the expression for $J_{1,2}$ in (\ref{Jlinearizable}).

The proof of the expression for $J_{3,4}$ is similar: Since the expression (\ref{S3S4explicit}) for $S_3(k)$ is analytic in $D_4$, the defining relation
\begin{align*}
J_{3,4} = e^{\hat{\mathcal{L}}x + \hat{\mathcal{Z}}t}(S_3^{-1}S_4),
\end{align*}
which a priori is valid only for $k \in \bar{D}_3 \cap \bar{D}_4$, can be extended to the line $k \in \bar{D}_{4} \cap \bar{D}_5$. 
After some algebra we find that $J_{3,4}$ depends on the entries of Ê$S(k)$ only via the four quotients
\begin{align}\label{J34quotients}
\frac{m_{13}(S)}{m_{33}(S)}, \quad 
\frac{m_{23}(S)}{m_{33}(S)},\quad
\frac{m_{11}(S)}{m_{31}(S)}, \quad  
\frac{m_{21}(S)}{m_{31}(S)}.
\end{align}
Solving the $(31)$ entry of (\ref{GRinfty}) and the equation obtained by letting $k \to -k$ in the $(32)$ entry of (\ref{GRinfty}) for the first two of these quotients, we find
\begin{subequations}\label{D5quotients}
\begin{align}
& \frac{m_{13}(S)}{m_{33}(S)} =
\frac{\hat{s}_{22} s_{31}-s_{21} \hat{s}_{32}}{\hat{s}_{12}
   s_{21}-s_{11} \hat{s}_{22}}, \qquad k \in D_5,
   	\\
&\frac{m_{23}(S)}{m_{33}(S)}= \frac{\hat{s}_{12} s_{31}-s_{11} \hat{s}_{32}}{\hat{s}_{12}
   s_{21}-s_{11} \hat{s}_{22}}, \qquad k \in D_5.
\end{align}
\end{subequations}
We use these relations to eliminate the first pair of quotients in (\ref{J34quotients}) from the expression for $J_{3,4}$ for $k \in \bar{D}_{4} \cap \bar{D}_5$.
We use the relations in (\ref{D3quotients}), which admit an analytic continuation to $D_4$, to eliminate the second pair of quotients in (\ref{J34quotients}). This yields the expression for $J_{3,4}$ given in (\ref{Jlinearizable}), but for $k \in \bar{D}_4 \cap \bar{D}_5$. However, since the expression for $J_{3,4}$ in (\ref{Jlinearizable}) is analytic in $D_4$, the equation is also valid for $k \in \bar{D}_3 \cap \bar{D}_4$.

Finally, the expression for $J_{4,5}$ follows simply by substituting the expressions for $S_4$ and $S_5$ in (\ref{Snexplicit}) into the definition
\begin{align*}
J_{4,5} = e^{\hat{\mathcal{L}}x + \hat{\mathcal{Z}}t}(S_4^{-1}S_5);
\end{align*}
in this case the $S(k)$ dependence disappears automatically.

We now show that the residue conditions in (\ref{Mres}) can be written as in (\ref{Mreslinearizable}).

In order to show (\ref{Mreslinearizablee}), we note that 
$$\frac{m_{21}(S(k_j))}{\dot{W}_3(k_j)} 
= \underset{k_j}{\res} \frac{m_{21}(S)}{W_3} 
= \underset{k_j}{\res} \frac{1}{\frac{m_{11}(S)}{m_{21}(S)}m_{22}(s) - m_{12}(s)},$$
where $k_j \in D_3$ is the zero of $W_3$.
Utilizing (\ref{D3quotients}) to eliminate the $S(k)$ dependence from this expression, we find after simplification
$$
\frac{m_{21}(S(k_j))}{\dot{W}_3(k_j)} 
= \underset{k_j}{\res} \frac{\hat{s}_{12} s_{33} - s_{13} \hat{s}_{32}}{s_{33}\mathcal{W}_4},$$
where $\mathcal{W}_4$ is given by (\ref{calW4def}).
Substitution of this into (\ref{Mrese}) yields (\ref{Mreslinearizablee}).

In order to show (\ref{Mreslinearizablef}), we use equations (\ref{D3quotients}) and (\ref{D5quotients}). As we have seen, both of these equations are valid for $k \in \bar{D}_4 \cap \bar{D}_5$. Algebraic manipulation using these relations shows that
$$\frac{S_{12}s_{23} - s_{13}S_{22}}{m_{32}(s)W_4}
= \frac{\hat{s}_{12} s_{23} - s_{13} \hat{s}_{22}}{m_{32}(s)\mathcal{W}_4}, 
\qquad  k \in \bar{D}_4 \cap \bar{D}_5,$$
and
$$\frac{S_{32}}{s_{33}W_4} = \frac{\hat{s}_{32}}{s_{33}\mathcal{W}_4}, \qquad k \in \bar{D}_4 \cap \bar{D}_5.$$
Since all quantities in these equations are analytic in $D_4$, these relations hold for all $k \in D_4$. In particular, they hold at $k_j \in D_4$. 
In view of (\ref{Mresf}), this yields (\ref{Mreslinearizablef}). 
The proofs of (\ref{Mreslinearizableb}) and (\ref{Mreslinearizablec}) are similar.
\proofend

\begin{remark}\label{linearizableremark}\upshape
It is tempting to attempt to derive the expressions for the matrices $J_{1,2}$ and $J_{3,4}$
in Theorem \ref{linearizableth} in the same way that the expression for $J_{2,3}$ was derived.
However, this does not appear to be possible. Indeed, consider the derivation of the expression for $J_{1,2} = e^{\hat{\mathcal{L}}x + \hat{\mathcal{Z}}t}(S_1^{-1}S_2)$.
In view of the expression (\ref{S2linearizable}) for $S_2(k)$, which is independent of $S(k)$, we only need to find an analogous expression for $S_1(k)$. 
From (\ref{S1S2explicit}) we see that $S_1(k)$ only depends on $S(k)$ via the quotients $S_{11}/S_{31}$ and Ê$S_{21}/S_{31}$. 
We can use the cofactor version of the global relation, which for $T = \infty$ is given by
\begin{align}\label{adGRinfty}
S^Ts^A = \begin{pmatrix}
m_{11}(c) & 0 & 0 \\
0 & m_{22}(c)  & 0 \\
0 & 0 & m_{33}(c)  
\end{pmatrix}, \qquad k \in \begin{pmatrix} D_{12} \cup D_1 & D_{6} & D_7  \\
D_{11} & D_{4} \cup D_{5} & D_{10}  \\
D_2 & D_3 & D_8 \cup D_9  \end{pmatrix},
\end{align}
to eliminate these quotients from $S_1(k)$. The $(11)$ entry of (\ref{adGRinfty}) and the equation obtained by letting $k \to -k$ in the $(13)$ entry of (\ref{adGRinfty}) provide two algebraic relations which can be solved for $S_{11}/S_{31}$ and Ê$S_{21}/S_{31}$ in $D_1$. The result is
\begin{align*}
& \frac{S_{11}}{S_{31}} = \frac{-m_{31}(s) m_{23}(\hat{s})+m_{21}(s) m_{33}(\hat{s})+ \frac{m_{11}(c)}{S_{31}} m_{23}(\hat{s})}{m_{11}(s)
   m_{23}(\hat{s})-m_{21}(s) m_{13}(\hat{s})}, \qquad k \in D_1,
   	\\
& \frac{S_{21}}{S_{31}} = \frac{-m_{31}(s) m_{13}(\hat{s})+m_{11}(s) m_{33}(\hat{s})+ \frac{m_{11}(c)}{S_{31}} m_{13}(\hat{s})}{m_{11}(s)
   m_{23}(\hat{s})-m_{21}(s) m_{13}(\hat{s})}, \qquad k \in D_1.
\end{align*}
We can use these relations to eliminate $S_{11}/S_{31}$ andÊ $S_{21}/S_{31}$ from $S_1(k)$. However, the result still involves the unknown quantity $\frac{m_{11}(c)}{S_{31}}$, hence the solution is not effective.
\end{remark}

\section{Non-linearizable boundary conditions}\nequation\label{nonlinearizablesec}
A major difficulty of initial-boundary value problems is that some of the boundary values are unknown for a well-posed problem. On the other hand, {\it all} boundary values are needed for the definition of $S(k)$, and hence for the formulation of the RH problem. In the previous section we analyzed the special case of linearizable boundary conditions---these are boundary conditions for which the matrix $S(k)$ can be eliminated algebraically from the formulation of the RH problem. In general, for non-linearizable boundary conditions, we need to determine $S(k)$ from the initial data and the given boundary values. In this section we concentrate on the effective characterization of $S(k)$. Following \cite{FLnonlinearizable}, we call a characterization of $S(k)$ {\it effective} if it fulfills the following requirements: (i) In the linear limit, it yields an effective solution of the linearized boundary value problem. (ii) For `small' boundary conditions, it yields an effective perturbative scheme, i.e. it yields an expression in which each term can be computed uniquely in a well-defined recursive scheme.

Our main result expresses the spectral function $S(k)$ in terms of the prescribed boundary data and the initial data via the solution of a system of nonlinear integral equations.
As in \cite{FLnonlinearizable}, our approach uses three ingredients: (a) The large $k$ asymptotics of the eigenfunctions. (b) The global relation. (c) A perturbative scheme to establish effectiveness.

\subsection{The global relation}
Evaluating (\ref{mu3mu2mu1sS}) at $(x,t) = (0,t)$, we find
$$\mu_2(0,t,k) e^{\hat{\mathcal{Z}}t}s(k) = c(t,k), \qquad k \in (\omega \mathcal{S}, \omega^2 \mathcal{S},\mathcal{S}),$$
where $c(t,k) = \mu_3(0,t,k)$. Defining functions $\{\Phi_j(t,k)\}_1^3$ and $\{c_j(t,k)\}_1^3$ by
\begin{align*}
\mu_2(0,t,k)
=
 \begin{pmatrix} \Phi_3(t,\omega^2 k) &  \Phi_2(t,\omega k) &  \Phi_1(t,k) \\
 \Phi_1(t,\omega^2 k) &  \Phi_3(t,\omega k) &\Phi_2(t,k) \\
 \Phi_2(t,\omega^2 k) &  \Phi_1(t,\omega k) &\Phi_3(t,k) \end{pmatrix}, \qquad
\frac{[c(t,k)]_3}{s_{33}(k)} = \begin{pmatrix}  c_1(t,k) \\
c_2(t,k) \\
c_3(t,k) \end{pmatrix}.
\end{align*}
we can write the $(13)$ and $(23)$ entries of this equation as 
\begin{subequations}\label{GR1GR2}
\begin{align}\label{GR1}
\Phi_3(t, \omega^2k)e^{(z_1 - z_3) t}\frac{s_{13}}{s_{33}} 
+ \Phi_2(t,\omega k)e^{(z_2 - z_3) t}\frac{s_{23}}{s_{33}}
+ \Phi_1(t,k) = c_1(t, k), \qquad k \in \mathcal{S},
	\\ \label{GR2}
\Phi_1(t,\omega^2k)e^{(z_1 - z_3) t}\frac{s_{13}}{s_{33}} 
+ \Phi_3(t,\omega k)e^{(z_2 - z_3) t}\frac{s_{23}}{s_{33}}
+ \Phi_2(t,k) = c_2(t, k), \qquad k \in \mathcal{S}.
\end{align}
\end{subequations}
The functions $\{c_j(t, k)\}_1^3$ are analytic and bounded in $\mathcal{S}$ away from the possible zeros of $s_{33}(k)$ and of order $O(1/k)$ as $k \to \infty$ in $\mathcal{S}$. The functions $\{\Phi_j(t, k)\}_1^3$ are entire functions of $k$ which are bounded forÊ $k \in \mathcal{T} \cup \hat{\mathcal{T}}$. 
Equation (\ref{GR1}) shows that whereas each of the functions $\Phi_3(t, \omega^2k)e^{(z_1 - z_3) t}$, 
$\Phi_2(t,\omega k)e^{(z_2 - z_3) t}$, and $\Phi_1(t,k)$ is bounded in $\mathcal{T} \cup \hat{\mathcal{T}}$, the combination appearing on the left-hand side of (\ref{GR1}) is bounded also in $\mathcal{S}$ (away from the zeros of $s_{33}$). A similar remark applies to (\ref{GR2}).

\subsection{Asymptotics}\label{asymptoticssubsec}
An analysis of (\ref{mulax}) shows that the eigenfunctions $\{\mu_j\}_1^3$ have the following asymptotics as $k \to \infty$: 
\begin{align}\label{mujnearzero}
&\mu_j(x,t,k) = I + \left[\biggl(\int_{(x_j, t_j)}^{(x,t)} \Delta\biggr)J^2
-
\frac{1}{i\sqrt{3}} \begin{pmatrix} 
 0 & -\omega ^2 q & \omega  r \\
 \omega ^2 r & 0 & -q \\
 -\omega  q & r & 0\end{pmatrix}
\right]\frac{1}{k}
	\\ \nonumber
& + \left[
\biggl(\int_{(x_j, t_j)}^{(x,t)} \eta_j \biggr)J
 + \frac{1}{3}\begin{pmatrix} 0 & \omega ^2 r^2 & \omega  q^2 \\
 q^2 & 0 & \omega  r^2 \\
 r^2 & \omega ^2 q^2 & 0\end{pmatrix}
 - \frac{i}{\sqrt{3}}
\biggl(\int_{(x_j, t_j)}^{(x,t)} \Delta\biggr)
 \begin{pmatrix} 
  0 & q & -r \\
 -\omega ^2 r  & 0 & \omega^2 q 
   \\
 \omega  q  & -\omega r  & 0
 \end{pmatrix}
 \right.
 	\\ \nonumber
&\left. \qquad + \frac{1}{3}\begin{pmatrix}
 0 & \omega  q_x & \omega ^2 r_x \\
 \omega  r_x & 0 & q_x \\
 \omega ^2 q_x & r_x & 0
\end{pmatrix}\right] \frac{1}{k^2} + O\Bigl(\frac{1}{k^3}\Bigr), \qquad k \to \infty, \quad j = 1,2,3,
\end{align}
where each column is valid within its region of boundedness and the closed one-forms $\Delta$ andÊ $\{\eta_j\}_1^3$ are defined by
\begin{align*}
\Delta = &\; qr dx + \biggl[\frac{i \left(q_x r-q r_x\right)}{\sqrt{3}}-\frac{2}{3} \left(q^3+r^3\right)\biggr]dt,
	\\
\eta_j =&\; d\biggl[\frac{1}{2}\biggl(\int_{(x_j, t_j)}^{(x,t)} \Delta\biggr)^2 - \frac{qr}{6}\biggr] - \biggl[\frac{i \left(q_x r-q r_x\right)}{2\sqrt{3}}-\frac{1}{3} \left(q^3+r^3\right)\biggr]dx
	\\
& - \biggl[q^2 r^2 + \frac{1}{3}q_xr_x - \frac{1}{6}(rq_{xx} + qr_{xx})\biggr]dt, \qquad j = 1,2,3.
\end{align*}
Hence
\begin{align}\label{s3expansion}
& [s(k)]_3 = 
 \begin{pmatrix} 0 \\ 0 \\ 1 \end{pmatrix}
 + \left[\begin{pmatrix} 0 \\ 0 \\ \omega^2 \int_{(\infty, 0)}^{(0,0)} \Delta \end{pmatrix}
-
 \frac{1}{i\sqrt{3}} \begin{pmatrix} 
 \omega  r(0,0) \\
 -q(0,0) \\
 0\end{pmatrix}
\right]\frac{1}{k} + O\Bigl(\frac{1}{k^2}\Bigr),
	\\ \nonumber
& \hspace{11cm} k \to \infty, \quad k \in \mathcal{S},
\end{align}
and
\begin{subequations}\label{Phiasymptotics}
\begin{align}\label{Phi1expansion}
& \Phi_1(t, k) = \frac{\Phi_1^{(1)}(t)}{k} + \frac{\Phi_1^{(2)}(t)}{k^2} + O \Bigl(\frac{1}{k^3} \Bigr),  && k \to \infty, \quad
k \in \mathcal{T} \cup \hat{\mathcal{T}},
	\\
& \Phi_2(t, k) =  \frac{\Phi_2^{(1)}(t)}{k}  + \frac{\Phi_2^{(2)}(t)}{k^2}  + O \Bigl(\frac{1}{k^3} \Bigr),  && k \to \infty, \quad
k \in \mathcal{T} \cup \hat{\mathcal{T}},
	\\
& \Phi_3(t, k) = 1 + \frac{\Phi_3^{(1)}(t)}{k}  + O \Bigl(\frac{1}{k^2} \Bigr), && k \to \infty, \quad
k \in \mathcal{T} \cup \hat{\mathcal{T}},
\end{align}
\end{subequations}
where
\begin{align*}
& \Phi_1^{(1)}(t) = -\frac{\omega h_0(t)}{i\sqrt{3}}, \qquad
\Phi_1^{(2)}(t) = \frac{\omega g_0(t)^2}{3} + \frac{ih_0}{\sqrt{3}}\int_{(0,0)}^{(0,t)} \Delta  + \frac{\omega^2h_1(t)}{3}, 
	\\
& \Phi_2^{(1)}(t) = \frac{g_0(t)}{i\sqrt{3}}, \qquad
\Phi_2^{(2)}(t) = \frac{\omega h_0(t)^2}{3} - \frac{i\omega^2g_0}{\sqrt{3}}\int_{(0,0)}^{(0,t)} \Delta  + \frac{g_1(t)}{3}, 
	\\
& \Phi_3^{(1)}(t) = \omega^2 \int_{(0, 0)}^{(0,t)} \Delta.
\end{align*}
In particular, we find the following expressions for the boundary values:
\begin{subequations}
\begin{align}\label{g0h0fromcoefficients}
&  g_0 = i\sqrt{3} \Phi_2^{(1)}, \qquad h_0 = -i \omega^2 \sqrt{3} \Phi_1^{(1)}, 
	\\ \label{g1h1fromcoefficients}
& g_1 = 3\Phi_2^{(2)} - \omega h_0^2 + i\sqrt{3} g_0 \Phi_3^{(1)}, 
\qquad
h_1 = 3\omega\Phi_1^{(2)} - \omega^2 g_0^2 - i\sqrt{3}\omega^2 h_0\Phi_3^{(1)}.
\end{align}
\end{subequations}

We will also need the asymptotics of $c_1$ and Ê$c_2$.
\begin{lemma}
The global relation (\ref{GR1GR2}) implies that the functions $c_1$ andÊ $c_2$ satisfy
\begin{subequations}
\begin{align}\label{c1asymptotics}
& c_1(t, k) = \frac{\Phi_1^{(1)}(t)}{k} + \frac{\Phi_1^{(2)}(t)}{k^2} + O \Bigl(\frac{1}{k^3} \Bigr), \qquad k \to \infty, \quad k \in \mathcal{S},
	\\ \label{c2asymptotics}
& c_2(t, k) = \frac{\Phi_2^{(1)}(t)}{k} + \frac{\Phi_2^{(2)}(t)}{k^2} + O \Bigl(\frac{1}{k^3} \Bigr), 
\qquad k \to \infty, \quad k \in \mathcal{S}.
\end{align}
\end{subequations}
\end{lemma}
\proofbegin
It follows from (\ref{mulax}) that the $\Phi_j$'s admit an expansion of the form (see chapter 6 of \cite{CL1955})
\begin{align} \label{Phialphabetagamma}
\begin{pmatrix}Ê\Phi_1(t,k) \\ \Phi_2(t,k) \\ \Phi_3(t,k) \end{pmatrix}
=&\; \Bigl(\alpha_0(t) + \frac{\alpha_1(t)}{k} + \cdots\Bigr)
+ \Bigl(\beta_0(t) + \frac{\beta_1(t)}{k} + \cdots\Bigr)e^{(z_1 - z_3)t}
	\\ \nonumber
& + \Bigl(\gamma_0(t) + \frac{\gamma_1(t)}{k} + \cdots\Bigr)e^{(z_2 - z_3)t}, \qquad k \to \infty, \quad k \in \C,
\end{align}
where the coefficients $\alpha_j(t)$, $\beta_j(t)$, $\gamma_j(t)$, $j = 0, 1, \dots$, are column vectors which are independent of $k$. We determine the coefficients by substituting (\ref{Phialphabetagamma}) into the system
\begin{align}  \nonumber
& \Phi_{1t} + (z_3 - z_1) \Phi_1 = \biggl(k \omega  g_0+\frac{i g_1}{\sqrt{3}}-h_0^2\biggr)\Phi_2
 + \biggl(k \omega ^2 h_0-g_0^2-\frac{i h_1}{\sqrt{3}}\biggl) \Phi_3,
	\\ \nonumber
& \Phi_{2t} + (z_3 - z_2)\Phi_2 = \biggl(k \omega  h_0-g_0^2-\frac{i h_1}{\sqrt{3}}\biggr)\Phi_1 
	+ \biggl(k g_0+\frac{i g_1}{\sqrt{3}}-h_0^2\biggr) \Phi_3,
	\\ \nonumber
& \Phi_{3t} = \biggl(k \omega ^2 g_0+\frac{i g_1}{\sqrt{3}}-h_0^2\biggr)\Phi_1 
 + \biggl(k h_0-g_0^2-\frac{i h_1}{\sqrt{3}}\biggr) \Phi_2,
\end{align}
and using the initial conditions
$$\alpha_0(0) + \beta_0(0) + \gamma_0(0) = (0,0,1)^T, \qquad
\alpha_1(0) + \beta_1(0) + \gamma_1(0) = (0,0,0)^T.$$
This yields
\begin{align*}
\begin{pmatrix}Ê\Phi_1(t,k) \\ \Phi_2(t,k) \\ \Phi_3(t,k) \end{pmatrix}
 &= \begin{pmatrix} 0 \\ 0 \\ 1 \end{pmatrix} + \begin{pmatrix} \Phi_1^{(1)}(t) \\ \Phi_2^{(1)}(t) \\  \Phi_3^{(1)}(t) \end{pmatrix} \frac{1}{k} + O\Bigl(\frac{1}{k^2}\Bigr)
	\\
& + \left\{ \begin{pmatrix} \frac{\omega h_0(0)}{i\sqrt{3}} \\ 0 \\ 0 \end{pmatrix} \frac{1}{k} +  O\Bigl(\frac{1}{k^2}\Bigr) \right\}e^{(z_1(k) - z_3(k))t}
	\\
& + \left\{\begin{pmatrix} 0 \\ \frac{i g_0(0)}{\sqrt{3}} \\ 0 \end{pmatrix} \frac{1}{k} + O\Bigl(\frac{1}{k^2}\Bigr) \right\}e^{(z_2(k) - z_3(k))t}, \qquad k \to \infty, \; k \in \C.
\end{align*}
Substituting this expansion and the expansion (\ref{s3expansion}) of $[s]_3$ into (\ref{GR1}), the resulting left-hand side involves the exponentials $e^{(z_1 - z_3)t}$ and $e^{(z_2 - z_3)t}$. Since the global relation requires that the right-hand side is of order $O(1/k)$ as $k \to \infty$ in $\mathcal{S}$, the coefficients of these exponentials must vanish. The remaining terms yield (\ref{c1asymptotics}). The expansion (\ref{c2asymptotics}) follows in a similar way from (\ref{GR2}). 
\proofend

\subsection{The Dirichlet and Neumann problems}
The following theorem expresses the spectral functions $A(k)$ and $B(k)$ in terms of the prescribed boundary data and the initial data via the solution of a system of nonlinear integral equations. 

\begin{theorem}\label{th1}
Let $T < \infty$. Let $q_0(x)$, $x \geq 0$, be a function of Schwartz class. For the Dirichlet problem it is assumed that the functions $g_0(t)$ and $h_0(t)$, $0 \leq t < T$, have sufficient smoothness and are compatible with $q_0(x)$ at $x=t=0$. Similarly, for the Neumann problem it is assumed that the functionsÊ $g_1(t)$ and $h_1(t)$, $0 \leq t < T$, have sufficient smoothness and are compatible with $q_0(x)$ at $x=t=0$.
Suppose that $s_{33}(k)$ is free from zeros (see remark \ref{zerosremark} below for the case when $s_{33}(k)$ has zeros). 

Then the spectral function $S(k)$ is given by
\begin{align}\label{SABC}
S(k) = \begin{pmatrix} A(\omega^2 k) &  B(\omega k) &  C(k) \\
 C(\omega^2 k) &  A(\omega k) &B(k) \\
 B(\omega^2 k) &  C(\omega k) &A(k) \end{pmatrix},
 \end{align}
 where
\begin{subequations}\label{ABCexpressions}
\begin{align}
& A(k)  = \Phi_3(T, \omega^2 k) \Phi_3(T, \omega k) - \Phi_1(T, \omega^2 k) \Phi_2(T, \omega k),
	\\
& B(k) = -\left[\Phi_3(T, \omega^2 k) \Phi_2(T, k) - \Phi_1(T, \omega^2 k) \Phi_1(T, k)\right]
e^{(z_3(k) - z_2(k))T},
	\\
& C(k) = \left[\Phi_2(T, \omega k) \Phi_2(T, k) - \Phi_3(T, \omega k) \Phi_1(T, k)\right]
e^{(z_3(k) - z_1(k))T} ,
\end{align}
\end{subequations}
and the complex-valued functions $\{\Phi_j(t, k)\}_1^3$ satisfy the following system of integral equations:
\begin{subequations}\label{Phieqs}
\begin{align}  \nonumber
& \Phi_1(t,k) = \int_0^t e^{(z_1(k) - z_3(k))(t - t')}\biggl[\biggl(k \omega  g_0+\frac{i g_1}{\sqrt{3}}-h_0^2\biggr)\Phi_2
	\\
&\hspace{5.6cm} + \biggl(k \omega ^2 h_0-g_0^2-\frac{i h_1}{\sqrt{3}}\biggl) \Phi_3 \biggr](t',k) dt',
	\\ \nonumber
& \Phi_2(t,k) = \int_0^t e^{(z_2(k) - z_3(k))(t - t')}\biggl[\biggl(k \omega  h_0-g_0^2-\frac{i h_1}{\sqrt{3}}\biggr)\Phi_1 
	\\
&\hspace{5.6cm}  + \biggl(k g_0+\frac{i g_1}{\sqrt{3}}-h_0^2\biggr) \Phi_3 \biggr](t',k) dt',
	\\ 
&\Phi_3(t,k) =1+ \int_0^t \biggl[\biggl(k \omega ^2 g_0+\frac{i g_1}{\sqrt{3}}-h_0^2\biggr)\Phi_1 
+ \biggl(k h_0-g_0^2-\frac{i h_1}{\sqrt{3}}\biggr) \Phi_2 \biggr](t',k) dt',
\end{align}
\end{subequations}

For a function $f(k)$, let $f_+(k)$ and $f_-(k)$ denote the following even and odd combinations of $f(k)$:
\begin{align*}
 & f_+(k) = f(k) + f(-k), \qquad f_-(k) = f(k) - f(-k), \qquad k \in \C.
\end{align*}  

\begin{itemize}
\item[(a)] For the Dirichlet problem, the unknown Neumann boundary values $g_1(t)$ and $h_1(t)$ are given by
\begin{subequations}\label{g1h1expressions}
\begin{align} \nonumber
g_1(t) = \;& \frac{3}{2\pi i}\int_{\partial \hat{\mathcal{S}}} \biggl[k\Phi_{2-}(t,k) - \frac{2 g_0(t)}{i\sqrt{3}}\biggr]dk - \omega h_0^2(t) + \frac{3\sqrt{3} g_0(t)}{2\pi}\int_{\partial \hat{\mathcal{S}}} \Phi_{3-}(t,k) dk
	\\ \label{g1expression}
& - \frac{3}{\pi i} \int_{\partial \hat{\mathcal{S}}} k\biggl[\Phi_1(t,-\omega^2k)e^{(z_1 - z_3) t}\frac{\hat{s}_{13}}{\hat{s}_{33}} 
+ \Phi_3(t,-\omega k)e^{(z_2 - z_3) t}\frac{\hat{s}_{23}}{\hat{s}_{33}}\biggr] dk, 
	\\ \nonumber
h_1(t) = \;& \frac{3\omega}{2\pi i}\int_{\partial \hat{\mathcal{S}}} \biggl[k\Phi_{1-}(t,k) + \frac{2\omega h_0(t)}{i\sqrt{3}}\biggr]dk
- \omega^2g_0^2(t)
- \frac{3\sqrt{3}\omega^2 h_0(t)}{2\pi}\int_{\partial \hat{\mathcal{S}}} \Phi_{3-}(t,k)dk
	\\ \label{h1expression}
& -\frac{3\omega}{\pi i} \int_{\partial \hat{\mathcal{S}}}k \biggl[\Phi_3(t, -\omega^2k) e^{(z_1 - z_3)t}\frac{\hat{s}_{13}}{\hat{s}_{33}}
+ \Phi_2(t,-\omega k) e^{(z_2 - z_3)t}\frac{\hat{s}_{23}}{\hat{s}_{33}}\biggr]dk.
 \end{align}
\end{subequations}

\item[(b)] For the Neumann problem, the unknown Dirichlet boundary values $g_0(t)$ and $h_0(t)$ are given by
\begin{subequations}\label{g0h0expressions}
\begin{align}\label{g0expression}  
  g_0(t) =\;& \frac{\sqrt{3}}{2\pi}\int_{\partial \hat{\mathcal{S}}} \Phi_{2+}(t,k) dk
  	\\\nonumber
&  - \frac{\sqrt{3}}{\pi} \int_{\partial \hat{\mathcal{S}}} \biggl[\Phi_1(t,-\omega^2k)e^{(z_1 - z_3) t}\frac{\hat{s}_{13}}{\hat{s}_{33}} 
+ \Phi_3(t,-\omega k)e^{(z_2 - z_3) t}\frac{\hat{s}_{23}}{\hat{s}_{33}}\biggr] dk,
	\\ \label{h0expression}
  h_0(t) = \; & -\frac{\omega^2 \sqrt{3}}{2\pi}\int_{\partial \hat{\mathcal{S}}} \Phi_{2+}(t,k) dk
  	\\\nonumber
& + \frac{\omega^2 \sqrt{3}}{\pi} \int_{\partial \hat{\mathcal{S}}} \biggl[\Phi_3(t, -\omega^2k) e^{(z_1 - z_3)t}\frac{\hat{s}_{13}}{\hat{s}_{33}}
+ \Phi_2(-\omega k) e^{(z_2 - z_3)t}\frac{\hat{s}_{23}}{\hat{s}_{33}}\biggr] dk.
\end{align}
\end{subequations}
\end{itemize}
\end{theorem}
\proofbegin
The representations (\ref{ABCexpressions}) follow from the relation $S(k) = e^{-\hat{\mathcal{Z}}T} \mu_2^A(0,T,k)^T$.

(a) In order to derive (\ref{g1h1expressions}) we note that equation (\ref{g1h1fromcoefficients}) expresses $g_1$ in terms of $\Phi_3^{(1)}$ and $\Phi_2^{(2)}$. Furthermore, equations (\ref{Phiasymptotics}) and Cauchy's theorem imply
\begin{align}\label{Phi31fromPhi3}
  -\frac{i\pi}{3} \Phi_3^{(1)}(t) = \int_{\partial \mathcal{T}} [\Phi_3(t,k) -1] dk = \int_{\partial \hat{\mathcal{T}}} [\Phi_3(t,k) - 1] dk
\end{align}
and
\begin{align}\label{Phi22fromPhi2}
  -\frac{i\pi}{3} \Phi_2^{(2)}(t) 
  = \int_{\partial \mathcal{T}} \biggl[k\Phi_2(t,k) - \Phi_2^{(1)}(t)\biggr] dk 
  = \int_{\partial \hat{\mathcal{T}}} \biggl[k\Phi_2(t,k) - \Phi_2^{(1)}(t)\biggr] dk.
\end{align}
Thus,
\begin{align}\nonumber
  \frac{2i\pi}{3} \Phi_3^{(1)}(t) & = -\biggl(\int_{\partial \mathcal{T}} + \int_{\partial \hat{\mathcal{T}}}\biggr) [\Phi_3(t,k) -1] dk
  = \biggl(\int_{\partial \hat{\mathcal{S}}} + \int_{\partial \mathcal{S}}\biggr) [\Phi_3(t,k) -1] dk
  	\\ \label{ipiPhi31}
&  = \int_{\partial \hat{\mathcal{S}}} [\Phi_3(t,k) -1] dk - \int_{\partial \hat{\mathcal{S}}} [\Phi_3(t,-k) -1] dk 
    = \int_{\partial \hat{\mathcal{S}}} \Phi_{3-}(t,k) dk
\end{align}
and
\begin{align}\nonumber
 \frac{2 i\pi}{3} \Phi_2^{(2)}(t) & = \biggl(\int_{\partial \hat{\mathcal{S}}} + \int_{\partial \mathcal{S}}\biggr) \bigl[k\Phi_2(t,k) - \Phi_2^{(1)}(t)\bigr] dk
  	\\ \nonumber
&  =  \biggl(\int_{\partial \hat{\mathcal{S}}} - \int_{\partial \mathcal{S}}\biggr) \biggl[k\Phi_2(t,k) - \Phi_2^{(1)}(t)\biggr] dk + I(t)
	\\ \label{ipiPhi22}
&  = \int_{\partial \hat{\mathcal{S}}}\bigl[k \Phi_{2-}(t,k) -  2\Phi_2^{(1)}(t)\bigr] dk 
       + I(t),
\end{align}
where $I(t)$Ê is defined by
$$I(t) = 2\int_{\partial \mathcal{S}} \bigl[k\Phi_2(t,k) - \Phi_2^{(1)}(t)\bigr] dk.$$
The last step involves using the global relation (\ref{GR2}) to compute $I(t)$:
\begin{align}\label{laststep}
I(t)
  = &\; 2\int_{\partial \mathcal{S}} \bigl[kc_2(t,k) - \Phi_2^{(1)}(t)\bigr] dk
  	\\\nonumber
&  - 2\int_{\partial \mathcal{S}} k\biggl[\Phi_1(t,\omega^2k)e^{(z_1 - z_3) t}\frac{s_{13}}{s_{33}} 
+ \Phi_3(t,\omega k)e^{(z_2 - z_3) t}\frac{s_{23}}{s_{33}}\biggr] dk
\end{align}
Hence, using Cauchy's theorem and the asymptotics (\ref{c2asymptotics}) of $c_2$ to compute the first term on the right-hand side of (\ref{laststep}) and using the transformation $k \to -k$ in the second term on the right-hand side of (\ref{laststep}), we find
\begin{align}  \nonumber
I(t) =& - \frac{4\pi i}{3} \Phi_2^{(2)}(t) 
 	\\ \label{secondtermcomputed}
& - 2\int_{\partial \hat{\mathcal{S}}} k\biggl[\Phi_1(t,-\omega^2k)e^{(z_1 - z_3) t}\frac{\hat{s}_{13}}{\hat{s}_{33}} 
+ \Phi_3(t,-\omega k)e^{(z_2 - z_3) t}\frac{\hat{s}_{23}}{\hat{s}_{33}}\biggr] dk.
\end{align}
Using (\ref{ipiPhi31}), (\ref{ipiPhi22}), and (\ref{secondtermcomputed}) in (\ref{g1h1fromcoefficients}) we find (\ref{g1expression}). The proof of (\ref{h1expression}) uses (\ref{GR1}) and is similar.

(b) In order to derive (\ref{g0h0expressions}) we note that equation (\ref{g0h0fromcoefficients}) expresses $g_0$ in terms of $\Phi_2^{(1)}$. Furthermore, equations (\ref{Phiasymptotics}) and Cauchy's theorem imply
\begin{align}\label{Phi21fromPhi2}
  -\frac{i\pi}{3} \Phi_2^{(1)}(t) = \int_{\partial \mathcal{T}} \Phi_2(t,k) dk = \int_{\partial \hat{\mathcal{T}}} \Phi_2(t,k) dk.
\end{align}
Thus,
\begin{align}\nonumber
  \frac{2i\pi}{3} \Phi_2^{(1)}(t) & = -\biggl(\int_{\partial \mathcal{T}} + \int_{\partial \hat{\mathcal{T}}}\biggr) \Phi_2(t,k)dk
  = \biggl(\int_{\partial \hat{\mathcal{S}}} + \int_{\partial \mathcal{S}}\biggr) \Phi_2(t,k)dk
  	\\ \label{ipiPhi21}
&  = \biggl(\int_{\partial \hat{\mathcal{S}}} - \int_{\partial \mathcal{S}}\biggr) \Phi_2(t,k) dk + K(t)
	\\ 
&  = \int_{\partial \hat{\mathcal{S}}} \Phi_{2+}(t,k) dk + K(t),
\end{align}
where $K(t)$Ê is defined by
$$K(t) = 2\int_{\partial \mathcal{S}} \Phi_2(t,k)dk.$$
The last step involves using the global relation (\ref{GR2}) to compute $K(t)$:
\begin{align}\label{Nlaststep}
K(t)
  = &\; 2\int_{\partial \mathcal{S}} c_2(t,k) dk
  	\\\nonumber
&  - 2\int_{\partial \mathcal{S}} \biggl[\Phi_1(t,\omega^2k)e^{(z_1 - z_3) t}\frac{s_{13}}{s_{33}} 
+ \Phi_3(t,\omega k)e^{(z_2 - z_3) t}\frac{s_{23}}{s_{33}}\biggr] dk.
\end{align}
Hence, using Cauchy's theorem and the asymptotics (\ref{c2asymptotics}) of $c_2$ to compute the first term on the right-hand side of (\ref{Nlaststep}) and using the transformation $k \to -k$ in the second term on the right-hand side of (\ref{Nlaststep}), we find
\begin{align}  \nonumber
K(t) =& - \frac{4\pi i}{3} \Phi_2^{(1)}(t) 
 	\\ \label{Nsecondtermcomputed}
& - 2\int_{\partial \hat{\mathcal{S}}} \biggl[\Phi_1(t,-\omega^2k)e^{(z_1 - z_3) t}\frac{\hat{s}_{13}}{\hat{s}_{33}} 
+ \Phi_3(t,-\omega k)e^{(z_2 - z_3) t}\frac{\hat{s}_{23}}{\hat{s}_{33}}\biggr] dk.
\end{align}
Using (\ref{ipiPhi21}) and (\ref{Nsecondtermcomputed}) in (\ref{g0h0fromcoefficients}) we find (\ref{g0expression}). The proof of (\ref{h0expression}) is similar.

\proofend

\begin{figure}
\begin{center}
\begin{overpic}[width=.44\textwidth]{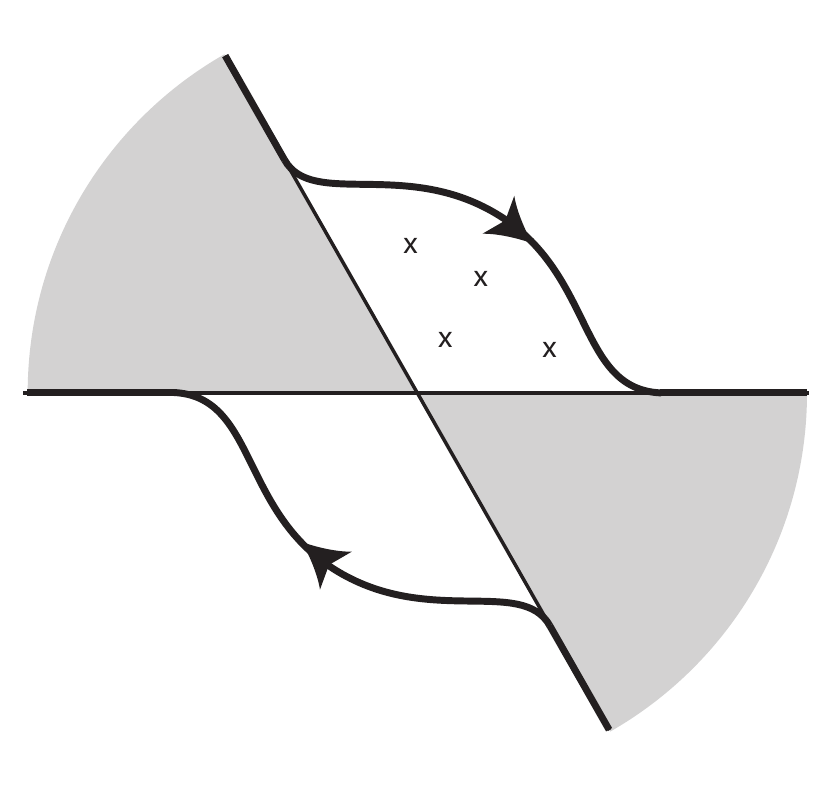}
      \put(20,63){$\mathcal{T}$}
      \put(75,29){$\hat{\mathcal{T}}$}
      \put(32,20){$\Gamma$}
      \put(60,73){$-\Gamma$}
       \put(101,47){$\text{Re}\; k$}
      \end{overpic}
     \begin{figuretext}\label{Gamma.pdf}
       The contour $\Gamma$ defined in such a way that $- \Gamma$ passes above the zeros of $s_{33}(k)$ in $\mathcal{S}$. The zeros of $s_{33}(k)$ are indicated by $x$'s in the figure.
     \end{figuretext}
     \end{center}
\end{figure}

\begin{remark}\upshape\label{zerosremark}
1. For the Dirichlet problem, substitution of the expressions (\ref{g1h1expressions}) for $g_1(t)$ and $h_1(t)$ into (\ref{Phieqs}) yields a system of quadratically nonlinear integral equations for the functions $\{\Phi_j(t, k)\}_1^3$. Similarly, for the Neumann problem, substitution of the expressions (\ref{g0h0expressions}) for $g_0(t)$ and $h_0(t)$ into (\ref{Phieqs}) yields a system of quadratically nonlinear integral equations for $\{\Phi_j(t, k)\}_1^3$.
Assuming that these systems have unique solutions, $S(k)$ can be determined from (\ref{SABC})-(\ref{ABCexpressions}). In fact, we will show in the following subsection that these nonlinear systems provide {\it effective} characterizations of $S(k)$ in the sense that they can be solved recursively to all orders in a perturbative scheme. 

2. In the case that $s_{33}(k)$ has a finite number of simple zeros in $\mathcal{S}$, the representations in (\ref{g1h1expressions}) and (\ref{g0h0expressions}) are valid provided that the integration contour $\partial \hat{\mathcal{S}}$ is replaced with $\Gamma$ everywhere, where $\Gamma$ is the contour obtained by deforming $\partial \hat{\mathcal{S}}$ in such a way that $-\Gamma$ passes above all the zeros of $s_{33}(k)$ in $\mathcal{S}$, see figure \ref{Gamma.pdf}.
\end{remark}

\subsection{Effective characterizations}\label{effectivesubsec}
Substituting into the system (\ref{Phieqs}) the expansions
\begin{align}
 \Phi_j &= \Phi_{j0} + \epsilon \Phi_{j1} + \epsilon^2 \Phi_{j2} + \cdots, \qquad  j = 1,2,3,
  	\\\label{g0g1expansions}
 g_0 &= \epsilon g_{01} + \epsilon^2 g_{02} + \cdots, \qquad  g_1 = \epsilon g_{11} + \epsilon^2 g_{12} + \cdots, 
	\\
 h_0 &= \epsilon h_{01} + \epsilon^2 h_{02} + \cdots, \qquad  h_1 = \epsilon h_{11} + \epsilon^2 h_{12} + \cdots, 	
\end{align}
where $\epsilon > 0$ is a small parameter, we find that the terms of $O(1)$ give $\Phi_{10} \equiv \Phi_{20} \equiv 0$ and $\Phi_{30} \equiv 1$. Moreover, the terms of $O(\epsilon)$ give $\Phi_{31} \equiv 0$ and
\begin{subequations}\label{eps1}
\begin{align}\label{eps1a}
&O(\epsilon): \;  \Phi_{11}(t,k) = \int_0^t e^{(z_1 - z_3)(t - t')} 
\biggl(k \omega^2 h_{01} - \frac{i h_{11}}{\sqrt{3}}\biggr) dt',
	\\
&O(\epsilon): \; \Phi_{21}(t,k) = \int_0^t e^{(z_2 - z_3)(t - t')}
\biggl(k g_{01}+\frac{i g_{11}}{\sqrt{3}}\biggr)  dt'.
\end{align}
\end{subequations}

On the other hand, expanding (\ref{g1h1expressions}) we find
\begin{subequations}\label{g11h11expressions}
\begin{align} 
g_{11} = \;& \frac{3}{2\pi i}\int_{\partial \hat{\mathcal{S}}} \biggl[ k\Phi_{21-} - \frac{2 g_{01}}{i\sqrt{3}}\biggr]dk - \frac{3}{\pi i}\int_{\partial \hat{\mathcal{S}}} k e^{(z_2 - z_3)t} \hat{s}_{231}dk,
	\\
h_{11} = \;& \frac{3\omega}{2\pi i}\int_{\partial \hat{\mathcal{S}}} \biggl[ k\Phi_{11-} + \frac{2 \omega h_{01}}{i\sqrt{3}}\biggr]dk - \frac{3\omega}{\pi i}\int_{\partial \hat{\mathcal{S}}} k e^{(z_1 - z_3)t} \hat{s}_{131}dk,	
\end{align}
\end{subequations}
where $s_{23} = \epsilon s_{231} + O(\epsilon^2)$ and $s_{13} = \epsilon s_{131} + O(\epsilon^2)$.
The Dirichlet problem can now be solved perturbatively as follows.
The odd parts of (\ref{eps1}) yield
\begin{align*}
& \Phi_{11-}(t,k) = 2\omega^2 k\int_0^t e^{(z_1 - z_3)(t - t')} h_{01} dt',
	\qquad \Phi_{21-}(t,k) = 2k\int_0^t e^{(z_2 - z_3)(t - t')}g_{01}  dt'.
\end{align*}
Given $g_{01}$ and $h_{01}$, we can use these equations to determine $\Phi_{11-}$ and $\Phi_{21-}$. We can then compute $g_{11}$ and $h_{11}$ from (\ref{g11h11expressions}) and then $\Phi_{11}$ and $\Phi_{21}$ follow from (\ref{eps1}). 
This recursive scheme can be continued indefinitely. Indeed, suppose $\Phi_{1j}$, $\Phi_{2j}$, $\Phi_{3j}$, $g_{1j}$, $h_{1j}$ have been determined for all $0 \leq j \leq n-1$ for some $n \geq 0$. The terms in (\ref{Phieqs}) of $O(\epsilon^n)$ give
\begin{align} \nonumber
&O(\epsilon^n): \; \Phi_{1n}(t, k) = \int_0^t e^{(z_1 - z_3)(t - t')} 
\biggl(k \omega^2 h_{0n} -\frac{i h_{1n}}{\sqrt{3}}\biggr) dt' + \lot,
	\\ \nonumber
&O(\epsilon^n): \; \Phi_{2n}(t, k) = \int_0^t e^{(z_2 - z_3)(t - t')}
\biggl(k g_{0n}+\frac{i g_{1n}}{\sqrt{3}}\biggr)  dt' + \lot,
	\\  \label{epsn}
&O(\epsilon^n): \; \Phi_{3n}(t, k) = \lot,
\end{align}
where `$\lot$' denotes an expression involving known terms of lower order. Similarly, the terms of $O(\epsilon^n)$ of (\ref{g1h1expressions}) give 
\begin{align}\label{g1nh1nexpressions}
g_{1n} = \;& \frac{3}{2\pi i}\int_{\partial \hat{\mathcal{S}}} \biggl[ k\Phi_{2n-} - \frac{2 g_{0n}}{i\sqrt{3}}\biggr]dk - \frac{3}{\pi i}\int_{\partial \hat{\mathcal{S}}} k e^{(z_2 - z_3)t} \hat{s}_{23n}dk+ \lot,
	\\ \nonumber
h_{1n} = \;& \frac{3\omega}{2\pi i}\int_{\partial \hat{\mathcal{S}}} \biggl[ k\Phi_{1n-} + \frac{2 \omega h_{0n}}{i\sqrt{3}}\biggr]dk - \frac{3\omega}{\pi i}\int_{\partial \hat{\mathcal{S}}} k e^{(z_1 - z_3)t} \hat{s}_{13n}dk  + \lot.
\end{align}
We find $\Phi_{1n-}$ and $\Phi_{2n-}$ from the odd parts of (\ref{epsn}); then $\{g_{1n}, h_{1n}\}$ follow from (\ref{g1nh1nexpressions}) and $\{\Phi_{jn}\}_{j=1}^3$ are finally found from (\ref{epsn}).
This shows that for the Dirichlet problem $\{\Phi_j\}_1^3$ can be determined to all orders in perturbation theory by solving the nonlinear system of Theorem \ref{th1} recursively.

Similarly, for the Neumann problem, the even parts of (\ref{eps1}) yield
\begin{align} \nonumber
&\Phi_{11+}(t,k) = -\frac{2i}{\sqrt{3}} \int_0^t e^{(z_1 - z_3)(t - t')} h_{11}(t') dt',
	\\ \label{Phi11Phi21plus}
&\Phi_{21+}(t,k) = \frac{2i}{\sqrt{3}} \int_0^t e^{(z_2 - z_3)(t - t')} g_{11}(t') dt',
\end{align}
while (\ref{g0h0expressions}) yields
\begin{align}\nonumber
  g_{01} =\;& \frac{\sqrt{3}}{2\pi}\int_{\partial \hat{\mathcal{S}}} \Phi_{21+} dk - \frac{\sqrt{3}}{\pi} \int_{\partial \hat{\mathcal{S}}} e^{(z_2 - z_3) t} \hat{s}_{231}dk,
	\\ \label{g01h01expressions}  
  h_{01} = \; & -\frac{\omega^2 \sqrt{3}}{2\pi}\int_{\partial \hat{\mathcal{S}}} \Phi_{21+} dk
 + \frac{\omega^2 \sqrt{3}}{\pi} \int_{\partial \hat{\mathcal{S}}} e^{(z_1 - z_3)t} \hat{s}_{131}dk.
\end{align}
Since $g_1$ and $h_1$ are known, (\ref{Phi11Phi21plus}) can be solved for $\Phi_{11+}$ and $\Phi_{21+}$, and then $g_{01}$ can be determined from (\ref{g01h01expressions}). Once $g_{01}$ and $h_{01}$ have been found, $\Phi_{11}$ and $\Phi_{21}$ can be computed from (\ref{eps1}). Extending this procedure to higher orders as in the case of the Dirichlet problem, we can determine $\{\Phi_j\}_1^3$ to all orders.

\section{Conclusions}\nequation \label{conclusionssec}
The IST formalism can be used to solve, at least in principle, the Cauchy initial value problem for an integrable PDE. However, many physically important problems requires solving an IBV problem. 
In this paper, our goal has been to analyze IBV problems for integrable evolution equations with $3 \times 3$ Lax pairs. To be concrete, we have considered the half-line problem for the system (\ref{qrsystem}) whose Lax pair can be viewed as a $3 \times 3$ generalization of the Lax pair for NLS. We have shown that: (a) The solution can be expressed in terms of the solution of a $3\times 3$-matrix RH problem in the complex $k$-plane formulated in terms of the spectral functions $s(k)$ and $S(k)$. Whereas $s(k)$ is defined in terms of the initial data alone, the definition of $S(k)$ requires knowledge of {\it all} the boundary values. (b) For the {\it linearizable} boundary conditions (\ref{linearizableBCs}), the function $S(k)$ can be eliminated by using only algebraic manipulation of the so-called global relation; the method is therefore as effective as the IST on the line. (c) In the general case of non-linearizable boundary conditions, the function $S(k)$ can be characterized in terms of the known boundary conditions via a system of nonlinear integrable equations. This characterization is effective in the sense that for small initial data it can be solved in a well-defined perturbative scheme.

For equations with Lax pairs involving $2 \times 2$ matrices, a method announced by Fokas in \cite{F1997, F2002} and subsequently developed by several authors has been successful in analyzing IBV problems. The results of this paper extend these ideas to equations with $3 \times 3$ Lax pairs. In fact, even though the transition to $3 \times 3$ Lax pairs requires a number of new steps, the end-result is that the main ingredients of the Fokas method can be implemented also for the system (\ref{qrsystem}). 
We expect that our approach will be applicable also to other equations with $3 \times 3$ Lax pairs.

\appendix
\section{Comparison with the case of $2 \times 2$-matrix Lax pairs} \label{Aapp}
\renewcommand{\theequation}{A.\arabic{equation}}\nequation
In this appendix we compare the constructions of this paper with the corresponding formalism for $2 \times 2$-matrix Lax pairs introduced by Fokas, see \cite{Fbook}. 
According to the approach of \cite{Fbook}, the analysis of an integrable PDE on the half-line with a $2Ê\times 2$ Lax pair of the form
\begin{equation}\label{mulaxapp}
\begin{cases}
  \mu_x + if_1(k) [\sigma_3, \mu] = V_1 \mu, \\
  \mu_t + i f_2(k) [\sigma_3, \mu] = V_2 \mu,
\end{cases} \qquad \sigma_3 = \begin{pmatrix} 1 & 0 \\ 0 & -1 \end{pmatrix},
\end{equation}
proceeds by defining eigenfunctions $\{\mu_j\}_1^3$ as solutions of the Volterra integral equations
\begin{equation}\label{mujdefapp}
  \mu_j(x,t,k) = I +  \int_{\gamma_j} e^{-i\theta(x,t,k) \hat{\sigma}_3} W_j(x',t',k), \qquad j = 1, 2,3,
\end{equation}
where $\theta(x,t,k) = f_1(k) x + f_2(k) t$, $W_j$ denotes the closed one-form
\begin{equation}\label{Wdefapp}
  W = e^{i\theta \hat{\sigma}_3}(V_1 dx + V_2 dt) \mu
\end{equation} 
with $\mu$ replaced by $\mu_j$, and the contours $\{\gamma_j\}_1^3$ are as in Figure \ref{mucontoursfig}. 
Then, defining sets $D_j \subset \C$, $j = 1,\dots, 4$, by
\begin{align} \nonumber
  D_1 = \{k | \text{Im}f_1 > 0 \; \text{and} \; \text{Im}f_2 > 0\},
  	\\ \nonumber
D_2 = \{k | \text{Im}f_1 > 0 \; \text{and} \; \text{Im}f_2 < 0\},
	\\ \label{Dnsapp}
D_3 = \{k | \text{Im}f_1 < 0 \; \text{and} \; \text{Im}f_2 > 0\},
	\\ \nonumber
D_4 = \{k | \text{Im}f_1 < 0 \; \text{and} \; \text{Im}f_2 < 0\},
\end{align}
a RH problem is formulated in terms of the sectionally analytic function $M$ defined by
\begin{align} \nonumber
& M = M_n, \qquad k \in D_n, \quad n = 1,\dots, 4;
	\\  \label{Mndefapp}
&M_1 = \left(\frac{[\mu_2]_1}{a(k)}, [\mu_3]_2\right), \qquad
M_2 = \left(\frac{[\mu_1]_1}{d(k)}, [\mu_3]_2\right),
	\\ \nonumber
& M_3 = \left([\mu_3]_1, \frac{[\mu_1]_2}{\overline{d(\bar{k})}}\right), \qquad
M_4 = \left([\mu_3]_1, \frac{[\mu_2]_2}{\overline{a(\bar{k})}}\right),
\end{align}
where the spectral functions $a,b,A,B$ are defined by\footnote{We assume, for simplicity, that the symmetries of (\ref{mulaxapp}) imply that $\sigma_1 \overline{\mu_j(x,t, \bar{k})} \sigma_1 = \mu_j(x,t, k)$, $j = 1,2,3$.}
$$\mu_3(0,0,k) = \begin{pmatrix} \overline{a(\bar{k})} & b(k) \\ \overline{b(\bar{k})} & a(k) \end{pmatrix}, \qquad
\mu_1(0,0,k) = \begin{pmatrix} \overline{A(\bar{k})} & B(k) \\ \overline{B(\bar{k})} & A(k) \end{pmatrix},$$
and $d(k) \equiv (\mu_3(0,T,k))_{22} = a(k) \overline{A(\bar{k})} - b(k) \overline{B(\bar{k})}$.

The eigenfunctions $\{\mu_j\}_1^3$ defined in (\ref{mujdef}) are the $3 \times 3$ analogs of the $\mu_j$'s defined in (\ref{mujdefapp}). We claim that the eigenfunctions $\{M_n\}_1^{12}$ used in the main text are the $3 \times 3$ analogs of the $M_n$'s defined in (\ref{Mndefapp}). To see this, we consider the $2 \times 2$ analog of the integral equation (\ref{Mnintegraleq}). Defining $\{l_j, z_j\}_1^2$ by
$$l_1(k) = -l_2(k) = -if_1(k), \qquad z_1(k) = -z_2(k) = -if_2(k),$$
we can write the Lax pair (\ref{mulaxapp}) as
$$d\left(e^{-\hat{\mathcal{L}} y - \hat{\mathcal{Z}} t} \mu \right) = W,$$
where $\mathcal{L} = \text{diag}(l_1, l_2)$ and $\mathcal{Z} = \text{diag}(z_1, z_2)$.
The $D_n$'s in (\ref{Dnsapp}) are given by
\begin{align*} 
  D_1 = \{k | \re l_1 > \re l_2 \; \text{and} \; \re z_1 > \re z_2\},
  	\\
D_2 = \{k | \re l_1 > \re l_2 \; \text{and} \; \re z_1 < \re z_2\},
	\\
D_3 = \{k | \re l_1 < \re l_2 \; \text{and} \; \re z_1 > \re z_2\},
	\\
D_4 = \{k | \re l_1 < \re l_2 \; \text{and} \; \re z_1 < \re z_2\}.
\end{align*}
Consider the eigenfunctions $\{\tilde{M}_n\}_1^4$ defined by the following $2 \times 2$ analog of the integral equation (\ref{Mnintegraleq}):
\begin{equation}\label{Mnintegraleqapp}
(\tilde{M}_n)_{ij}(x,t,k) = \delta_{ij} + \int_{\gamma_{ij}^n} \left(e^{\hat{\mathcal{L}}(k)x + \hat{\mathcal{Z}}(k) t} W_n(x',t',k)\right)_{ij}, \qquad k \in D_n, \quad i,j = 1, 2,
\end{equation}
where the contours $\gamma^n_{ij}$, $n = 1, \dots, 4$, $i,j = 1, 2$, are defined by (\ref{gammaijnudef}) and $W_n$ is given by (\ref{Wdefapp}) with $\mu$ replaced with $\tilde{M}_n$.
The following lemma shows that the $\tilde{M}_n$'s defined by equation (\ref{Mnintegraleqapp}) coincide with the $M_n$'s defined in (\ref{Mndefapp}).

\begin{lemma}\label{Mnsequallemma}
  We have $\tilde{M}_n = M_n$, $n = 1, \dots, 4$.
\end{lemma}
\proofbegin
Define spectral functions $S_n(k)$, $n = 1, \dots, 4$, by
\begin{align*}
  S_n(k) = \tilde{M}_n(0,0,k), \qquad k \in D_n.
\end{align*}
Then
\begin{equation}\label{Mnmujrelationsapp}
  \tilde{M}_n = \mu_2 e^{\hat{\mathcal{L}} x + \hat{\mathcal{Z}} t} S_n, \qquad n = 1, \dots, 4.
\end{equation}  
Define $s(k)$ and $S(k)$ by $s(k) = \mu_3(0,0,k)$ and $S(k) = \mu_1(0,0,k)$. 

We will give a proof in the case of $n = 1$; similar arguments apply when $n = 2,3,4$.
The $2 \times 2$ matrix $(\gamma^1)_{ij} := \gamma^1_{ij}$ is given by
$$\gamma^1 = \begin{pmatrix} \gamma_3 & \gamma_3 \\ \gamma_2 & \gamma_3 \end{pmatrix}.$$
Proceeding as in the proof of Lemma \ref{Snexplicitlemma}, we find the following $2\times 2$ analog of the matrix factorization problem (\ref{sSSnrelations}):
$$  s(k; X_0) = S_1(k; X_0)T_1^{-1}(k; X_0), \qquad S(k) = S_1(k; X_0)R_1^{-1}(k; X_0), \qquad k \in D_1,$$
where
$$R_1(k; X_0) = \begin{pmatrix} * & * \\ * & * \end{pmatrix}, \qquad
S_1(k; X_0) = \begin{pmatrix} * & * \\ 0 & * \end{pmatrix}, \qquad
T_1(k; X_0) = \begin{pmatrix} 1 & 0 \\ * & 1 \end{pmatrix}$$
and $*$ denotes an entry yet to be determined.
These are eight equations for eight unknowns. Solving these equations and letting $X_0 \to \infty$, we find
\begin{align*}
& S_1(k) = \begin{pmatrix} \frac{1}{a(k)} & b(k) \\ 
0& a(k)\end{pmatrix}.
\end{align*}
Consequently, by (\ref{Mnmujrelationsapp}),
$$\tilde{M}_1 = \mu_2 e^{\hat{\mathcal{L}} x + \hat{\mathcal{Z}}t} S_1
= \left(\frac{[\mu_2]_1}{a}, b e^{-2i\theta} [\mu_2]_1 + a[\mu_2]_2 \right).$$
Since the $2 \times 2$ analog of equation (\ref{mu3mu2mu1sS}) shows that 
$$b e^{-2i\theta} [\mu_2]_1 + a[\mu_2]_2 = [\mu_3]_2,$$
this expression for $\tilde{M}_1$ indeed coincides with the expression for $M_1$ given in (\ref{Mndefapp}). 

\proofend

\begin{remark}\upshape
In view of lemma \ref{Mnsequallemma}, equation (\ref{Mnintegraleqapp}) provides an alternative definition of the $M_n$'s. In the case of $2 \times 2$ matrices, since in each domain $D_n$, $n = 1, \dots, 4$, there exist two column vectors of the $\mu_j$'s which are bounded and analytic, the $M_n$'s are conveniently defined by combining the column vectors of the $\mu_j$'s according to (\ref{Mndefapp}). However, in the case of $3 \times 3$ matrices, the $M_n$'s involve rather complicated combinations of the entries of the $\mu_j$'s, and because of the relatively small domains of boundedness of the $\mu_j$'s, the boundedness properties of these combinations are not evident. We were therefore instead led to generalize the alternative definition (\ref{Mnintegraleqapp}).
\end{remark}

We can also verify directly that the $2 \times 2$ analog of our definition of the jump matrices $J_{m,n}$ reproduces the jump conditions used in \cite{Fbook}. Indeed, the $2 \times 2$ analog of definition (\ref{Jmndef}) is
$$J_{m,n} =
e^{\hat{\mathcal{L}} x + \hat{\mathcal{Z}}t}(S_m^{-1}S_n),\qquad n, m = 1, \dots, 4.$$
For definiteness, we consider the jump across $\bar{D}_1 \cap \bar{D}_2$. Arguments similar to those used in the proof of Lemma \ref{Mnsequallemma} show that
$$S_2(k) = \begin{pmatrix} \frac{\overline{A(\bar{k})}}{d(k)} & b(k) \\ 
\frac{\overline{B(\bar{k})}}{d(k)} & a(k)\end{pmatrix}.$$
Thus, the jump condition across $\bar{D}_1 \cap \bar{D}_2$ is given by
$$M_2 = M_1 J_{1,2}, \qquad k \in \bar{D}_1 \cap \bar{D}_2\quad \text{where} \quad 
J_{1,2} = \begin{pmatrix} 1 & 0 \\ \frac{\overline{B(\bar{k})}}{a(k) d(k)} e^{2 i \theta} & 1 \end{pmatrix},$$
which is the jump condition given in \cite{Fbook}.

\section{Fredholm integral equations}\nequation\label{fredholmapp}
\renewcommand{\theequation}{B.\arabic{equation}}\nequation
We will show that an extension of the standard Fredholm theory can be used to establish the required existence and analyticity properties of the solutions $\{M_n\}_1^{12}$ of the integral equations (\ref{Mnintegraleq}). This will complete the proof of Proposition \ref{Mnprop}. 

Fix $n \in \{1, \dots, 12\}$ and $j \in \{1,2,3\}$. Letting $w_i(x,t,k) = (M_n)_{ij}(x,t,k)$, we can write the $j$'th column of (\ref{Mnintegraleq}) as
\begin{equation}\label{wieq}
w_i(x,t,k) = \delta_{ij} + \int_{\gamma_{ij}^n} \sum_{l = 1}^3 K_j(x,t; x',t'; k)_{il} w_l(x',t',k), \quad i = 1, 2,3,
\end{equation}
where the kernel $K_j$ is defined by
\begin{align}\label{Kjkernel}
K_j(x,t;x',t';k)_{il} = & e^{(l_i - l_j)(x - x') + (z_i - z_j)(t - t')}
	\\ \nonumber
& \times (V_1(x',t',k)dx' + V_2(x',t',k)dt')_{il}, \qquad i, l = 1,2,3.
\end{align}
There are two reasons why the standard Fredholm theory does not immediately apply to (\ref{wieq}): (a) The equation involves the three different integration contours $\{\gamma_{j}\}_1^3$ instead of being defined on (part of) the real line. (b) The integral kernel $K_j$ is, in general, not an $L^2$-kernel. Indeed, for $k$ such that $ \re  l_i(k) =  \re  l_j(k)$, the exponential factor in (\ref{Kjkernel}) is bounded, but does not decay as $x,x' \to \infty$. In this case, the kernel therefore has decay as $x' \to \infty$, but not necessarily as $x \to \infty$.

In order to address (a), we first consider the equation on the boundary
\begin{equation}\label{boundary}  
  \{(x,0)\, | \, 0 < x < \infty\} \cup \{(0, t)\, | \, 0 < t < T\}.
\end{equation}
The idea is to map this boundary to the semi-infinite interval $[-T, \infty)$. Define $\theta(x)$ by $\theta(x) = 1$ for $x \geq 0$ and $\theta(x) = 0$ for $x < 0$. 
We define a $3 \times 3$-matrix valued kernel $\mathcal{K}_j(s,s', k)$ for $s,s' \in (-T, \infty)$ as follows. Let $K_j = K_j^{(x)} dx + K_j^{(t)} dt$. For $s > 0$, we let
\begin{align}
&  \mathcal{K}_j(s,s', k) = \begin{cases} -K_j^{(t)}(s,0; 0, |s'|; k), &  \gamma_{ij}^n = \gamma_1 \ \text{and} \ -T < s' < 0, \\
\theta(s - s')K_j^{(x)}(s,0; s', 0; k), &  \gamma_{ij}^n = \gamma_{1,2}  \ \text{and} \  0 < s' < \infty, \\
0,  & \gamma_{ij}^n = \gamma_2  \ \text{and} \  -T < s' < 0, \\
-\theta(s' - s)K_j^{(x)}(s,0; s', 0; k), & \gamma_{ij}^n = \gamma_3  \ \text{and} \  -T < s' < \infty,
\end{cases}
\end{align}
while, for $-T < s < 0$, we let
\begin{align}
&  \mathcal{K}_j(s,s', k) = \begin{cases} -\theta(s - s') K_j^{(t)}(0, |s|; 0, |s'|; k), &  \gamma_{ij}^n = \gamma_1  \ \text{and} \  -T < s' < \infty, \\
\theta(s' - s)K_j^{(t)}(0, |s|; 0, |s'|; k), &  \gamma_{ij}^n = \gamma_{2, 3} \ \text{and} \  -T < s' < 0, \\
0,  & \gamma_{ij}^n = \gamma_2  \ \text{and} \  0 < s' < \infty, \\
-K_j^{(x)}(0, |s|; s', 0; k), & \gamma_{ij}^n = \gamma_3  \ \text{and} \  0 < s' < \infty.
\end{cases}
\end{align}
Then, introducing the vector valued function $v(s,k)$ by
$$v_i(s,k) = \begin{cases} w_i(0, |s|,k), & -T < s < 0, \\ 
w_i(s,0, k), & 0 < s < \infty,
\end{cases}$$
we can write (\ref{wieq}) as
\begin{equation}\label{vieq}  
  v_i(s,k) = \delta_{ij} + \int_{-T}^\infty  \sum_{l = 1}^3 \mathcal{K}_j(s,s',k)_{il}v_l(s',k) ds', \qquad -T < s < \infty, \quad i = 1,2,3.
\end{equation}  
This is a Fredholm equation of the second kind. It was observed by Caudrey \cite{C1982} that although the kernel $\mathcal{K}_j$ is, in general, not of $L^2$-type, equations such as (\ref{vieq}) can be analyzed by an extension of the standard Fredholm theory, cf. \cite{H1973, T1957}. Following \cite{C1982}, we find that a sufficient condition on the kernel $\mathcal{K}_j$ for equation (\ref{vieq}) to be solvable is that there exists a function $m(s', k) > 0$ such that
$$|\mathcal{K}_j(s,s',k)_{il}| < m(s', k), \qquad i,l = 1, 2,3, \quad s, s' \in (-T, \infty),$$
and
$$\int_{-T}^\infty m(s', k)ds' = \mathcal{M}(k),$$
where $\mathcal{M}(k)$ is a bounded function of $k$. If we consider (\ref{vieq}) with $k \in \bar{D}_n$, $n =1, \dots, 12$, the boundedness of the exponential factor in (\ref{Kjkernel}) for $k \in \bar{D}_n$ implies that such an $\mathcal{M}(k)$ exists. We infer that the Fredholm determinant $f(k)$ and minor $F(s,s',k)$ associated with (\ref{vieq}) are defined and analytic in the interior of the set $D_n$ and are continuous up to the boundary of $D_n$. Provided that $f(k) \neq 0$ the solution of (\ref{vieq}) is given by
$$v_i(s,k) = \delta_{ij} + \int_{-T}^\infty \frac{F(s,s',k)_{ij}}{f(k)} ds'.$$
The zeros of $f(k)$ are the eigenvalues of the Fredholm equation for $v_i$. Since $f(k)$ is analytic this set is discrete. This shows that $M_n(x,t,k)$ has the properties stated in Proposition \ref{Mnprop} whenever $(x,t)$ belongs to the boundary (\ref{boundary}).

Assuming now that the values of $w_i(x,t,k)$ for $(x,t)$ on the boundary (\ref{boundary}) are known, we can proceed to define $M_n(x,t,k)$ also for $(x,t)$ in the set $\{0 < x < \infty, 0 < t < T\}$. Define the vector valued function $w_0(t, k)$ by
$$w_{0i}(t, k) = \begin{cases}
\delta_{ij} + \int_T^t K_j^{(t)}(x,t; 0,t'; k)_{il}w_l(0,t',k) dt', & \gamma_{ij}^n = \gamma_1, \\
\delta_{ij} + \int_0^t K_j^{(t)}(x,t; 0,t'; k)_{il}w_l(0,t',k) dt', & \gamma_{ij}^n = \gamma_2, \\
\delta_{ij}, &  \gamma_{ij}^n = \gamma_3, \\
\end{cases} \qquad i = 1,2,3.
$$
Then (\ref{wieq}) can be written as
\begin{equation}\label{wiw0eq}
w_i(x,t,k) = w_0(t,k) + \int_0^\infty \sum_{l = 1}^3 \tilde{K}_j(x, x', t; k)_{il} w_l(x',t,k) dx', \quad i = 1, 2,3,
\end{equation}
where
$$\tilde{K}_j(x, x', t; k)_{il} = \begin{cases} \theta(x-x')K_j^{(x)}(x,t; x', t; k), &  \gamma_{ij}^n = \gamma_{1,2}, \\
- \theta(x' - x)K_j^{(x)}(x,t; x', t; k), &  \gamma_{ij}^n = \gamma_3.
 \end{cases}$$
For each fixed $t \in (0,T)$, equation (\ref{wiw0eq}) is a Fredholm equation to which we may apply the approach of \cite{C1982}. It follows that the solution $M_n(x,t,k)$ exists for all $(x,t)$ and has the properties stated in Proposition \ref{Mnprop}.

\bigskip
\noindent
{\bf Acknowledgement} {\it The author thanks the referee for helpful remarks and acknowledges support from the EPSRC, UK.}

\bibliographystyle{plain}
\bibliography{is}

\begin{thebibliography}{99}
\small

\bibitem{BC1984}
R. Beals and R. R. Coifman, Scattering and inverse scattering for first order systems, {\it Comm. Pure Appl. Math.} {\bf 37} (1984), 39--90.

\bibitem{BFS2006}
A. Boutet De Monvel, A. S. Fokas, and D. Shepelsky, Integrable nonlinear evolution equations on a finite interval,
{\it Comm. Math. Phys.} {\bf 263} (2006), 133--172.

\bibitem{BS1998}
A. Boutet de Monvel and D. Shepelsky, Inverse scattering problem for a stratified bi-isotropic medium at oblique incidence, {\it Inverse Problems} {\bf 14} (1998), 29--40.

\bibitem{BS2000}
A. Boutet de Monvel and D. Shepelsky, A frequency-domain inverse problem for a dispersive stratified chiral medium, {\it J. Math. Phys.} {\bf 41} (2000), 6116--6129.

\bibitem{BS2003}
A. Boutet de Monvel and D. Shepelsky, The modified KdV equation on a finite interval,
{\it C. R. Math. Acad. Sci. Paris} {\bf 337} (2003), 517--522.

\bibitem{BS2004}
A. Boutet de Monvel and D. Shepelsky, Initial boundary value problem for the mKdV equation on a finite interval,
{\it Ann. Inst. Fourier (Grenoble)} {\bf 54} (2004), 1477--1495.

\bibitem{C1982}
P. J. Caudrey, The inverse problem for a general $N \times N$ spectral equation,
{\it Physica D} {\bf 6} (1982), 51--66.

\bibitem{CL1955}
E. A. Coddington and N. Levinson, {\it Theory of differential equations}, New York, McGraw-Hill, 1955.

\bibitem{DP1999}
A. Degasperis and M. Procesi, Asymptotic integrability,
in {\it Symmetry and Perturbation Theory}, edited by A. Degasperis and G.
Gaeta, World Scientific (1999), pp. 23--37.

\bibitem{DTT1982}
P. Deift, C. Tomei, and E. Trubowitz, Inverse scattering and the Boussinesq equation, {\it Comm. Pure Appl. Math.} {\bf 35} (1982), 567--628. 

\bibitem{DZ1993}
P. Deift and X. Zhou, A steepest descent method for oscillatory Riemann-
Hilbert problems. Asymptotics for the MKdV equation, 
{\it Ann. of Math.} {\bf 137} (1993), 295--368.

\bibitem{F1997}
A. S. Fokas, A unified transform method for solving linear and certain nonlinear PDEs, 
{\it Proc. Roy. Soc. Lond.} A {\bf 453} (1997), 1411--1443.

\bibitem{F2002}
A. S. Fokas, Integrable nonlinear evolution equations on the half-line, 
{\it Comm. Math. Phys.} {\bf 230} (2002), 1--39.

\bibitem{Fbook}
A. S. Fokas, A unified approach to boundary value problems, CBMS- NSF regional conference series in applied mathematics, SIAM (2008).

\bibitem{FI1994}
A. S. Fokas and A. R. Its, An initial-boundary value problem for the Korteweg-de Vries equation, 
{\it Math. Comput. Simulation} {\bf 37} (1994), 293--321.
 
 
\bibitem{FIS2005}
A. S. Fokas, A. R. Its, and L.-Y. Sung, The nonlinear Schr\"odinger equation on the half-line, 
{\it Nonlinearity} {\bf 18} (2005), 1771--1822.

\bibitem{FLnonlinearizable}
A. S. Fokas and J. Lenells, The unified method: I Non-linearizable problems on the half-line, preprint, arXiv:1109.4935.

\bibitem{H1973}
H. Hochstadt, {\it Integral equations}, Pure and Applied Mathematics, John Wiley \& Sons, New York-London-Sydney, 1973.

\bibitem{Kamvissis}
S. Kamvissis, Semiclassical nonlinear Schr\"odinger on the half line,
{\it J. Math. Phys.} {\bf 44} (2003), 5849--5868. 

\bibitem{K1980}
D. J. Kaup, On the inverse scattering problem for cubic eigenvalue problems of the class $\psi _{xxx}+6Q\psi _{x}+6R\psi =\lambda \psi $, 
{\it Stud. Appl. Math.} {\bf 62} (1980), 189--216. 

\bibitem{Lholedisk}
J. Lenells, Boundary value problems for the stationary axisymmetric Einstein equations: a disk rotating around a black hole, {\it Comm. Math. Phys.} {\bf 304} (2011), 585--635.

\bibitem{LFernst}
J. Lenells and A. S. Fokas, Boundary-value problems for the stationary axisymmetric Einstein equations: a rotating disc, {\it Nonlinearity} {\bf 24} (2011), 177--206.

\bibitem{M1974}
S. V. Manakov, On the theory of two-dimensional stationary self focussing of electromagnetic waves, {\it Sov. Phys. JETP} {\bf 38} (1974), 248--253.

\bibitem{M1981}
A. V. Mikhailov, The reduction problem and the inverse scattering method,
{\it Physica D} {\bf 3} (1981), 73--117.

\bibitem{MK2006}
E. A. Moskovchenko and V. P. Kotlyarov, A new Riemann-Hilbert problem in a model of stimulated Raman scattering,
{\it J. Phys. A} {\bf 39} (2006), 14591--14610. 


\bibitem{P2005}
B. Pelloni, The asymptotic behavior of the solution of boundary value problems for the sine-Gordon equation on a finite interval,
{\it J. Nonlinear Math. Phys.} {\bf 12} (2005), 518--529.

\bibitem{SS1991}
N. Sasa and J. Satsuma, New-type of soliton solutions for a higher-order nonlinear Schr\"odinger equation, {\it J. Phys. Soc. Japan} {\bf 60} (1991), 409--417.

\bibitem{SK1974}
K. Sawada and T. Kotera, A method for finding $N$-soliton solutions of the K.d.V. equation and K.d.V.-like equation, {\it Progr. Theoret. Phys.} {\bf 51} (1974), 1355--1367.

\bibitem{SS2000}
D. Sheen and D. Shepelsky, Uniqueness in a frequency-domain inverse problem of a stratified uniaxial bianisotropic medium, {\it Wave Motion} {\bf 31} (2000), 371--385.

\bibitem{T1957}
F. G. Tricomi, {\it Integral equations}, Pure and Applied Mathematics, Vol. V Interscience Publishers, New York; Interscience Publishers Ltd., London, 1957.

\bibitem{ZM1973}
V. E. Zakharov and S.V. Manakov, Resonant interaction of wave packets in nonlinear media, {\it JETP Lett.} {\bf 18} (1973), 243--245.


\end{thebibliography}

\end{document}